\documentclass[%
 aip,
 amsmath,amssymb,
reprint,
]{revtex4-1}

\usepackage{graphicx}
\usepackage{dcolumn}
\usepackage{bm}

\usepackage[utf8]{inputenc}
\usepackage[T1]{fontenc}
\usepackage{mathptmx}
\usepackage[T1]{fontenc} 
\usepackage{amsmath,color}
\usepackage{amssymb}
\usepackage{amsfonts}
\usepackage{amsthm}
\usepackage{mathtools}
\usepackage[title]{appendix}
\usepackage{hyperref}
\usepackage{braket}
\usepackage{xcolor}
\usepackage{etoolbox}
\usepackage{soul}

\newcommand*{\citen}[1]{%
  \begingroup
    \romannumeral-`\x 
    \setcitestyle{numbers}%
    \cite{#1}%
  \endgroup
}

\usepackage{algorithm}
\usepackage{algpseudocode}
\usepackage{dsfont}

\usepackage{bbold}

\renewcommand\thefigure{\arabic{figure}}
\begin{document}

\title{Heat Transport with a Twist}

\author{Ethan Abraham}
\email{abrahame@sas.upenn.edu}
\affiliation{Department of Physics and Astronomy, University of Pennsylvania, Philadelphia, Pennsylvania 19104, USA }
\author{Mohammadhasan Dinpajooh}
\email{hadi.dinpajooh@pnnl.gov, mdinpajo@sas.upenn.edu}
\affiliation{Physical and Computational Sciences Directorate, Pacific Northwest National Laboratory, WA 99352 USA}
\author{Clàudia Climent}
\affiliation{Department of Chemistry, University of Pennsylvania, Philadelphia, Pennsylvania 19104, USA }
\author{Abraham Nitzan}
\affiliation{Department of Chemistry, University of Pennsylvania, Philadelphia, Pennsylvania 19104, USA }
\affiliation{School of Chemistry, Tel Aviv University, Tel Aviv 69978, Israel}

\date{\today}

\begin{abstract}
\vspace{1 cm}
Despite the desirability of polymers for use in many products due to their flexibility, light weight, and durability, their status as thermal insulators has precluded their use in applications where thermal conductors are required. However, recent results suggest that the thermal conductance of polymers can be enhanced and that their heat transport behaviors may be highly sensitive to nanoscale control.
Here we use non-equilibrium molecular dynamics (MD) simulations to study the effect of mechanical twist on the steady-state thermal conductance across multi-stranded polyethylene wires. We find that a highly twisted double-helical polyethylene wire can display a thermal conductance up to three times that of its untwisted form, an effect which can be attributed to a structural transition in the strands of the double helix. We also find that in thicker wires composed of many parallel strands, adding just one twist can increase its thermal conductance by over $30\%.$ However, we find that unlike stretching a polymer wire, which causes a monotonic increase in thermal conductance, the effect of twist is highly non-monotonic, and certain amounts of twist can actually decrease the thermal conductance. Finally, we apply the Continuous Chirality Measure (CCM) in an attempt to explore the correlation between heat conductance and chirality. The CCM is found to correlate with twist as expected, but we attribute the observed heat transport behaviors to structural factors other than chirality.

\end{abstract}
\maketitle
	
	
\section{Introduction}
\label{intro}
For many years, the ability to control heat transport on a macroscopic scale has defined important technologies such as refrigeration devices, heating systems, and thermal insulators. One would also expect that the future ability to control heat transport on the nanoscale will play a pivotal role in the success of technologies such as nanoelectronics and nanoscale energy devices that require thermal management.\cite{general1} In this regard, polymers are materials that may be desirable for use in many such products due to their flexibility, light weight, and resistance to corrosion, yet their status as thermal insulators has precluded their use in products where thermal conductors like metals and ceramics are required.\cite{h2, general2, general3} This makes the nanoscale control and improvement of heat transport in polymers an important engineering application.\cite{h2}

It is now known that the heat transport properties of polymers are highly sensitive to external conformational factors. For example, while the thermal conductivity of polymers in bulk amorphous form is often quite small, these same polymers when aligned in crystalline form can display, in the chain direction, a thermal conductivity that begins to approach that of a conductor.\cite{Hadi, h4} Multiple recent studies have investigated and found effects of mechanical strain on the heat conduction in aligned polymer materials,\cite{h4, h1, h2, h3, h5, h6} and previous work in our group has found that stretching a thin polymer wire can increase its thermal conductivity by up to an order of magnitude.\cite{Hadi,Hadi-2} Another less explored conformational factor is twist. Controlled twisting has recently been used as a way to study chirality effects on molecular transport properties,\cite{Twist1, Twist2, Twist3} and in a recent study Wang \textit{et al.} have used twist to engineer fibers that cool either during stretch or stretch release.\cite{Twist-Cooling} This observation of twist-based cooling, which they attributed in the case of polyethylene to entropy changes associated with deformation-driven monoclinic-to-orthorhombic phase conversions, shows promise as a novel and potentially sustainable approach to cooling.\cite{Twist-Cooling} All of this invites one to explore the general effect of twist on the heat transport properties of molecular chains. 

While MD simulations have been used to compute the heat transport across a polymer wire as a function of stretch, the same has yet to be performed when the wire is twisted.\cite{Hadi, h5} In an analogous fashion to previous work in our group that controlled stretch by changing the end-to-end distance across a polyethylene wire of fixed size,\cite{Hadi} here we control the twist in such a wire. The effect of twisting a single polyethylene chain is null due to the invariance of the C-C sigma bonds under complete rotation, so the simplest model system for the present study is a two-stranded polyethylene wire, which forms the familiar double helix when twisted. In addition to its direct relevance for heat transport applications, this study is also motivated by recent theoretical interest in the way chirality is manifested in molecular and condensed phase optical and transport phenomena,\cite{n1, n2, n3, n4, n5, n6, n7, Spin-Seedback, Chiral_Heat_Transport} as exemplified by recent discussions of possible correlations between chirality and heat conduction in carbon nanotubes,\cite{Nanotubes1, Nanotubes2, Nanotubes3, Nanotubes4, Nanotubes5, Nanotubes6, Nanotubes7} and by recent observations that implicate nuclear motions of chiral molecules in the observed temperature dependence of some of these phenomena.\cite{Temp_Effects_CISS, Spin-Seedback, Nanotubes5}

We note that chirality, as a symmetry property, has traditionally been viewed as a binary classification of molecules as either chiral or achiral. Yet in the past three decades, quantification of chirality has been considered.\cite{c1, c1b, c2, c3, c4, c5, c6, Temp_CCM} Here and in a follow-up paper, we will show that the control parameter of twist in our system does correlate with these chiral observables, and so the model we develop in this paper provides a system with a continuous variation of chirality that can be used as a laboratory for studying chiral effects on molecular transport properties. 

It should be pointed out, as already suggested in Ref. \citen{Nanotubes5}, that the twist-dependence is not necessarily associated with the chiral nature of the twisted structures and may be better attributed to mechanical factors such as torsional strain or interactions between the chains. Like in many other systems, analysis of the heat transport mechanisms in polymer wires has revealed a competition between diffusive and ballistic transport mechanisms. Diffusive transport generally occurs when the length scale of molecular interactions is much smaller than the length of the transport process of interest and is stochastic in nature. On the other extreme, ballistic transport occurs by the free motion of phonons, and such dynamics are dominated by the harmonic part of the molecular force field (FF). Polymer wires are of such a size and connectivity that the relative contribution of each can be sensitive to external variables. For example, the study by Dinpajooh’s \textit{et al.} showed characteristics of mostly ballistic transport when the single polymer is stretched, and characteristics of diffusive transport when more slack is present.\cite{Hadi} Note that ballistic transport by delocalized phonons is free from the stochastic back-scattering events that characterize diffusive transport and is therefore associated with a higher thermal conductance. We have seen that stretching can increase the ballistic nature of the heat transport, and we will see here (\hyperref[results]{Sec. III}) that to some extent twisting has a similar effect, albeit in a more complex and non-monotonous fashion.

In \hyperref[details]{Sec. II}, we present the technical details of our simulations. In \hyperref[results]{Sec. III}, we discuss our results pertaining to the effect of twist on molecular heat conduction in polymer wires for different chain lengths, molecular force fields, and number of chains. We also discuss the possible correlation of these results with other physical observables such as chirality. We conclude in \hyperref[conclusion]{Sec. IV}.

\section{Methods}
\label{details}

{\it Molecular Model.} For the majority of our calculations, we have followed Dinpajooh \textit{et al.} in representing our polymers using the Transferable Potentials for Phase Equilibria (TraPPE) United Atom (UA) model, which treats each CH$_x$ unit as a single particle.\cite{Hadi, FF3, FF5} This model is useful because it greatly reduces the computational cost, yet it has been shown to still perform with high accuracy for calculations of thermal conductance in hydrocarbons.\cite{unitedWorks} It is based on a force field (FF) that represents bonds with harmonic potentials and includes also angle (3-body), dihedral (4-body), and Lennard Jones (LJ) potentials. The Hamiltonian is given by\begin{equation}
\begin{split}
H_{\rm{molecule}} & = \sum_i \frac{p_i^2}{2m_i} +  \sum_i k_{bi} (l_i-l_{0i})^2 + \sum_i k_{\theta i} (\theta_i-\theta_{0i})^2 + \\
  & \sum_i \sum_{n_i}^4 \frac{C_{n i}}{2} \left[ 1+ (-1)^{n{_i}-1} {\rm{cos}} ( n_i \phi_i) \right] + \\
  & \sum_i \sum_j 4 \epsilon_{ij} \left[  (\frac{\sigma_{ij}}{r_{ij}})^{12} - (\frac{\sigma_{ij}}{r_{ij}})^{6}  \right],
\end{split}
\label{Hmol}
\end{equation}where $p_i$ and $m_i$ are the momentum and mass of a given particle respectively, $l_i$,$l_{0i}$, $\theta_i$,$\theta_{0i}$, are the actual and equilibrium bond lengths and angles respectively, $\phi_i$ are the dihedral angles, and $r_{ij}$ are the inter-particle separations. Accordingly, the parameters $k_{bi}$ and $k_{\theta i}$ are the spring constants for the bonds and angles, and the $C_{ni}$ are the constants that define the dihedral potentials. As in Ref. \citen{Hadi}, the above parameter values were chosen to fit observed physical properties.

We have compared results with two other models: i) an Explicit Hydrogen (EH) model and ii) a simplified force field inspired by the Freely Joint Chain (FJC) model.\cite{FJC} The Explicit Hydrogen model uses the same Hamiltonian as above, except that the Hydrogen and Carbon atoms are treated as separate bodies and different parameter values are chosen as appropriate, which are available in the Supplementary Information  (SI). The FJC force field is also a united atom model that conventionally omits the angular potentials, dihedral potentials, and Leonard Jones potentials. In the present application, since our calculations pertain to wires with multiple strands, the omission of the LJ potentials would create a non-physical situation with no repulsive force between two chains. We, therefore, modify the FJC field to include the LJ potential and denote this force field as FJC$^*$. Hence the Hamiltonian becomes \begin{equation}
H_{\rm{FJC^*}} = \sum_i \frac{p_i^2}{2m_i} +  \sum_i k_{bi} (l_i-l_{0i})^2 + \sum_i \sum_j 4 \epsilon_{ij} \left[  (\frac{\sigma_{ij}}{r_{ij}})^{12} - (\frac{\sigma_{ij}}{r_{ij}})^{6}  \right],
\label{FRC}
\end{equation}using the same parameter values as the TraPPE-UA FF. It should be noted that this FJC$^*$ model allows the angles to relax from $\theta_0=114^{\circ}$ to $180^{\circ}$, changing the natural length of the polymer. Therefore we report two separate sets of results for the FJC$^*$ FF, one that keeps the same end-to-end distance as the previous calculations (FJC$^*$ Original Length), and one that is adjusted to the new natural length, which is the same as the contour length of the polymer (FJC$^*$ Contour Length).

\begin{figure}[tbh]
\includegraphics[width=1\columnwidth]{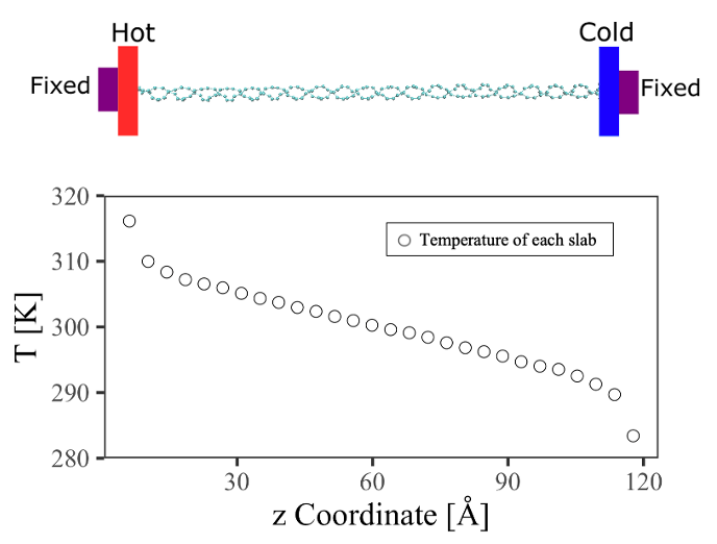}\caption{Top: a schematic of a double-helical polyethylene wire with 10 twists in a non-equilibrium MD simulation of steady state heat transport. The atoms in the slabs at each end are fixed to act as a wall, and those in the adjacent slabs are treated as Langevin thermostats, which set the imposed
temperature difference. Bottom: the steady-state temperature profile obtained for the same polymer when the left and right temperatures are set to TL = 320 K and TR = 280 K.}
\label{FIG1}
\end{figure}

{\it Calculation of Thermal Conductance.} Our simulation consists of a polymer wire held at either end by parallel walls of specified temperature (see \hyperref[FIG1]{Fig.  \ref{FIG1}}). This is obtained by designating, at each end of the wire, a segment of the chain (typically 3 atoms long) to be fixed, followed on either side by a segment (roughly $6$ \AA\ long) in which the atoms are set to move as Markovian Langevin thermostats. We note that the designation of internal atoms as thermostats is justified since we are concerned not with junction effects but with the intrinsic conduction of the polymer. The rest of the wire’s length is then divided into segments (typically 20-30 segments of equal length) for computation of the coarse-grained temperature along the wire. The (classical) time evolution of the system is computed according to the above-described Hamiltonian using LAMMPS (March 2018 version 16 with corrections for heat flux).\cite{LAAMPS1, LAAMPS2} The timestep used was 1.25 fs for the UA models and 0.65 fs for the Explicit Hydrogen model.\footnote{A shorter timestep is preferred for the Explicit Hydrogen model due to the higher frequencies of the C-H bond vibrations.} The steady state heat current is then computed as the average over the two sides of the energy taken out $(\Delta E_{\text{hot}})$ of the hot thermostat region and added $(\Delta E_{\text{cold}})$ to the cold thermostat region per unit time
\begin{equation}
\tag{3a}
I_z  = \frac{1}{2}\left(  \left|\frac{\Delta E_{\rm{hot}} }{\Delta t}\right|  + \left|\frac{\Delta E_{\rm{cold}} }{\Delta t}\right|     \right).
\end{equation}
The thermal conductance is then obtained from
\begin{equation}
\tag{3b}
\label{eq:conductance}
K = I_z /\Delta T.
\end{equation}
In addition, temperature profiles are obtained by monitoring the kinetic energy in each of the segments mentioned above to plot temperature as a function of position across the wire.

These heat transport simulations were then carried out as in Ref. \citen{Hadi, NEMD1} using the standard velocity-Verlet time integrator. The starting configurations were first equilibrated under canonical NVT (constant number of particles, volume, and temperature) for several nanoseconds using the Nose–Hoover thermostat at 300 K for all atoms.
The outputted equilibrated configurations were then used as the starting configurations for the non-equilibrium Molecular Dynamics (NEMD) simulations, in which Langevin thermostats were used only for the hot and cold regions. The target temperatures for the hot and cold regions were 320 K and 280 K.
We used a damping parameter of $50$ fs$^{-1}$ for the thermostated regions, and we observed that the results have minimal dependency on the damping parameters ranging from $10$ to $100$ fs$^{-1}$ (results not shown).
After relaxation to the steady state for about 1 ns, the simulations were continued for 40 ns to obtain data for statistical averaging.
Five independent simulations were used to further reduce statistical uncertainties. Note the heat transport behavior is reported in terms of heat conductance (Eq. \ref{eq:conductance}) because, as discussed in Ref. \citen{Hadi} and also below, heat transport on the nanoscale is not necessarily diffusive, making the intensive concept of conductivity difficult to define.

{\it Generation of Twisted Structures.} Twisted configurations were prepared by applying a torque to one end of the wire while keeping the other end fixed. The LAMMPS command \texttt{addtorque} was used to apply the torque. Untwisted wires sampled from equilibrium room temperature simulations were used as the starting configuration for these twist-generating dynamics. The number of twists was found to grow continuously from 0 at the start of the simulation until a point where bonds began to break as a result of the torque.\footnote{We found that for the purposes of generating these twisted structures, applying a large torque for a short time was more practical than applying a small torque and waiting for equilibrium.} The number of twists at any MD step was computed as half the number of crosses between the two chains in the wire\footnote{When twisting is done about the z axis, a “cross” is defined as a change of sign of $x_j - x_j'$ (or $y_j - y_j')$ as $j$ runs from $1$ to $N$ along the chain, where $x_j$ and $x_j'$ (or $y_j$ and $y_j'$) refer to the coordinates of atom $j$ on different chains.} and verified by visual inspection of the structures using VMD software.\cite{vmd} The coordinate files from the MD steps corresponding to the desired number of twists were then used as the starting configurations for the heat transport simulations described above. For further details and proof of principle regarding these twist-generating simulations, see SI. 

Note that a multi-strand polymer wire of a given length is characterized by the maximum number of twists that can be imposed on it before bonds start breaking. We infer that bonds are breaking when the end of the wire receiving the torque continues to rotate, yet the number of twists does not increase. We denote this maximal number of twists, which depends on the chain length $N$, by $N_T^{\text{max}}$, and we define $\lambda_{\text{twist}}(N) = N_T^{\text{max}}(N)/N$. Results below indicate that $N_T^{\text{max}}$ is linear in $N$ (at least to a very good approximation), so $\lambda_{\text{twist}}$ does not depend on $N$ and is a property of the wire's constitution.

{\it Other Calculations.}

In order to gain further insight into the relationship between the structures and the heat transport behavior, more numerical analysis was performed on the wire configurations. The normal modes and corresponding frequencies were obtained from energy minimized structures. Energies were minimized using a simple “steepest descent minimizer” such that the minimization was converged when the maximum force was smaller than $25$  kJ mol$^{-1}$  nm$^{-1}$), and normal modes were then obtained by diagonalizing the Hessian matrix. These calculations were carried out using GROMACS software.\cite{gromacs} A canonical localization measure for normal modes is the \textit{participation ratio ($P_k$)} of mode $k$, varying from $1$ when the vibrations are localized on a single atom to $N$ if the mode is equally distributed on all atoms. The participation ratio is computed as follows: let the coefficients $u^\alpha_{k,i}$ denote the expansion of normal mode $k$ in the atomic coordinates where $\alpha$ denotes the Cartesian coordinate and $i$ denotes the atom number. Defining $p_{k,i}=\sum_{\alpha} |u_{k,i}^\alpha|^2$, the participation ratio is then defined as \begin{equation}\tag{4}\label{eq:Pk}P_k = 1/\sum^N_i p^2_{k,i}.\end{equation} Note that since we have taken the modes to be normalized, we have $\sum_i \sum_\alpha |u^\alpha_{k,i}|^2 = \sum^N_i p_{k,i}=1$.\cite{participationRatio1,participationRatio2} 
We have also used the Savitzky–Golay filter method to filter the set of participation ratio data points $P_k$ for the purpose of smoothing the data,\cite{smooth} which in our experience is a way to avoid averaging over thousands of frames to get the participation ratios. The smoothing method involves fitting successive sub-sets of adjacent data points with a low-degree polynomial by the method of linear least squares, and we found that the difference between the average $P_k$ sum from several frames and the smoothed $P_k$ sum from one frame was less than $0.1\%$.

Finally, noting that the double-helical wire is a model chiral system, we have applied the Continuous Chirality Measure (CCM), a mathematical tool for measuring chirality that returns 0 for an achiral structure and 1 for a structure that is very different than its enantiomer.\footnote{In many of the mathematical texts where the CCM is developed, the CCM is said to range from 0 to 100. We as physical scientists, prefer our dimensionless quantities to be on the order of unity. The only difference between the two approaches is the factor of 100 which we omit.} For mathematical details of the CCM, see SI and Ref. \citen{c1b, c2}. These CCM calculations were implemented by the \texttt{gsym} function of the \texttt{cosymlib} library in Python.\cite{cosymlib} The CCM was computed using real-time configurations from our MD calculations rather than energy-minimized structures, raising the possibility that random thermal fluctuations could slightly affect the CCM. To account for this, the reported CCM values are thermal averages taken from at least five configurations.

\section{Results \& Discussion}
\label{results}

In studies of polymer wires, the simplest system to examine is typically a single chain of polyethylene. However, twisting across the sp$^3$ sigma bonds has no persistent effect on the structure, rendering such a model ineffective when the control is twist. Therefore, most of the results presented below take our model system to be a two-stranded polyethylene chain, which forms the familiar double helix when twisted. Studies of DNA usually define \textit{twist} as the total number of complete helical turns in a given DNA segment,\cite{DNAtwist} which is mathematically equivalent to Gauss's linking number evaluated between curves representing the two strands.\cite{Gauss} We extend this same definition to our polyethylene double helix and refer to it as the \textit{number of twists} ($N_T$) to avoid ambiguity. A left-handed helix has been used for all of our reported results, but we performed several trials to confirm the intuition that the right-handed helix exhibits the same thermal conductance behavior. Unless otherwise specified, we choose the end-to-end distance of the polymer wire to be its natural length, that is the maximal extension that can be obtained without stretching any bonds or angles away from their equilibrium values. In the case of polyethylene, this corresponds to a fraction $R_{\rm{frac}}=0.83$ of the chain contour length. Here $R_{\rm{frac}}=R_{\rm{ete}}/R_{\rm{contour}}$ where $R_{\rm{contour}}$ is the contour length of the polymer and $R_{\rm{ete}}$ is the given end-to-end distance. Note that the contour length as defined here is equivalent to the equilibrium bond length times the number of bonds.

\begin{figure}[tbh]
\includegraphics[width=1\columnwidth]{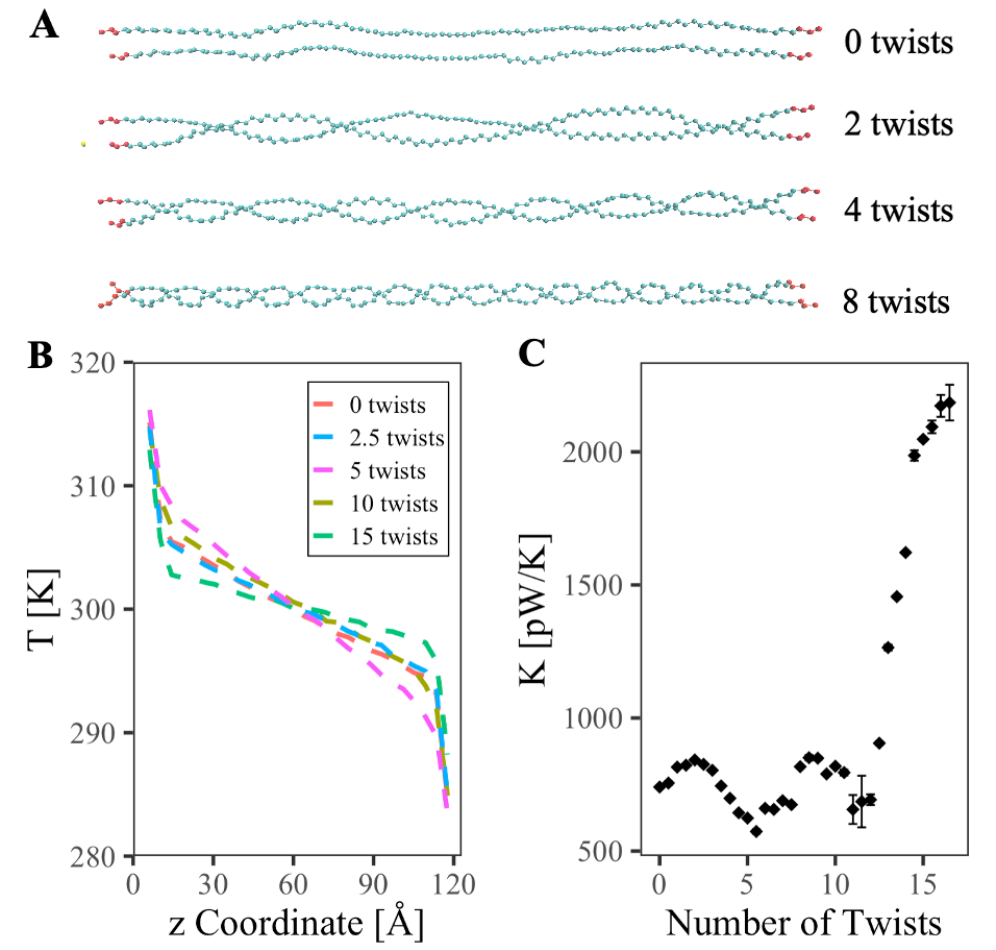}
\caption{\textbf{(A)} Cartoon visualizations of an $N=98$ double-helical polyethylene wire containing various numbers of twists. The atoms shown in red at the ends of the wires are fixed in position throughout the simulations. \textbf{(B)} The temperature profiles along the direction of the wire ($z$ coordinate) at selected levels of twist. \textbf{(C)} Thermal conductance results for the wire as a function of the number of twists. The statistical uncertainties (standard deviations) are shown when larger than the symbol sizes.}
\label{FIG2}
\end{figure}

In what follows, we report our results for heat conduction and temperature profiles of polymer wires subject to various amounts of twist. We also look for correlations between the observed heat conduction behaviors and other properties such as chirality, potential energy, and mode localization for the twisted structures. In these NEMD simulations, the thermostats operating at the two ends of the molecular chain were set to 280 K and 320 K. Unless otherwise specified, the TraPPE-UA force field has been used to model the polymers.

\hyperref[FIG2]{Figure  \ref{FIG2}A} shows cartoon representations of double helices composed of two $N=98$ polyethylene chains containing various levels of twists. The atoms marked in red are fixed throughout the simulations. \hyperref[FIG2]{Figure  \ref{FIG2}B} shows average temperature profiles for the NEMD simulations at five selected numbers of twists, and \hyperref[FIG2]{Fig.  \ref{FIG2}C} the results of our thermal conductance calculations. For this chain length, the maximal number of twists that could be obtained before bonds began to break was 17, so we refer to this number as the \textit{twist saturation} point and report results only for levels of twist less than or equal to this amount. The statistical uncertainties marked in \hyperref[FIG2]{Fig. \ref{FIG2}C} are obtained using five independent NEMD simulations of 40 ns each.

Apart from the quantitative thermal conductance results, there are several key qualitative observations. The most striking is that when the point of twist saturation is approached, the thermal conductance spikes to roughly three times the value of the untwisted wire. Also noteworthy is the oscillating pattern for lower levels of twist. The conductance increases as the number of twists increases from 0 to 2 twists at which point a relative maximum is obtained, but then decreases as the number of twists increases from 2 to 5.5, at which point a relative minimum is obtained. Another oscillation with a similar period and amplitude occurs between 5.5 and 11 twists before the final spike occurs. 
The temperature profiles show that for levels of twist where the conductance is high, the majority of the temperature bias falls on the edges whereas for levels of twist where the conductance is lower, we see more of the temperature bias spread out over the length of the chain. This is best seen by comparing the temperature profile for 15 twists (green line in \hyperref[FIG2]{Fig. \ref{FIG2}B}), which has a high conductance and shows temperature bias mostly near the interfaces, to that of 5 twists (purple line), which has a low conductance and thus a profile that falls more uniformly across the junction. As discussed in earlier work,\cite{Hadi} a large temperature drop at the interfaces and a flatter temperature profile in the interior of the junction indicates a more ballistic character that involves less arbitrary back-scattering events, making the heat transport more efficient.

\begin{figure}[tbh]
\includegraphics[width=1\columnwidth]{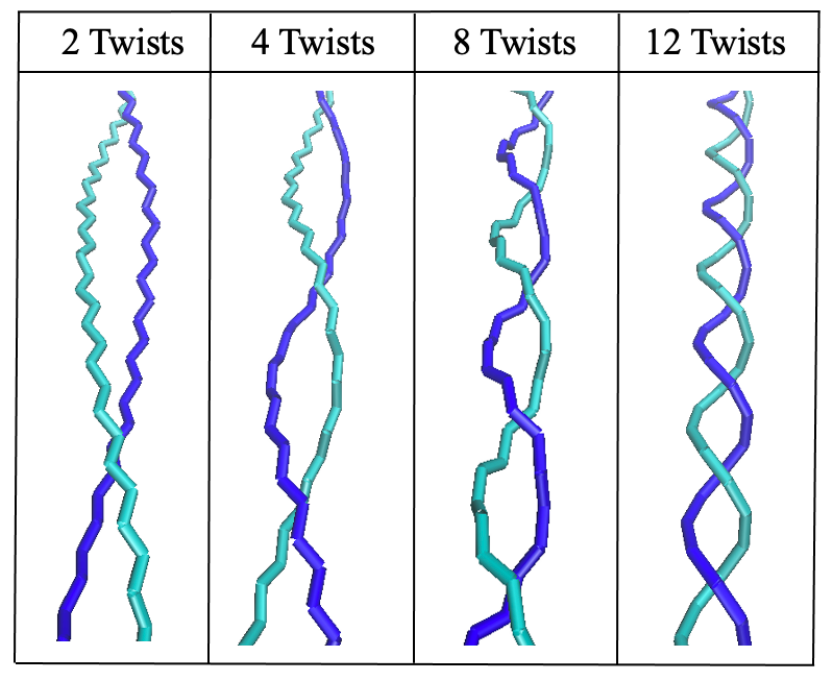}
\caption{Representation of the bonds in PE polymer structures for selected number of twists. Snapshots are shown (from left to right) for 2 twists, 4 twists, 8 twists, and 12 twists. The two chains are visually distinguished by different shades of blue.}
\label{FIG8}
\end{figure}

Though we could not generally pinpoint the microscopic origin of this difference in back-scattering behavior when the number of twists is varies, direct visualizations of PE polymer chains show features of the structures that change categorically where the high-twist spike begins. \hyperref[FIG8]{Figure \ref{FIG8}} shows representations of the bonds in our model double-helix with a selected number of twists. As can be seen, at low levels of twist, the bonds make a local zig-zag pattern that is expected for an ethylene chain. This persists even at 8 twists although the zig-zag pattern becomes a bit more complex. However, at 12 twists, the local structures almost completely transition to the helical curve. This categorical smoothing of the structure appears to coincide perfectly with the observed increase in thermal conductance, and it is therefore likely that the ultra-high conductance at high twist is related to a decrease in phonon back scattering associated with this structural transition. In SI, Fig. S4 shows that a similar structural transition can be induced by holding the number of twists constant and varying the LJ intermolecular interactions between the chains. This suggests that the interchain interactions are critical to the observed increase in thermal conductance at high twist.

\begin{figure}[tbh]
\includegraphics[width=1\columnwidth]{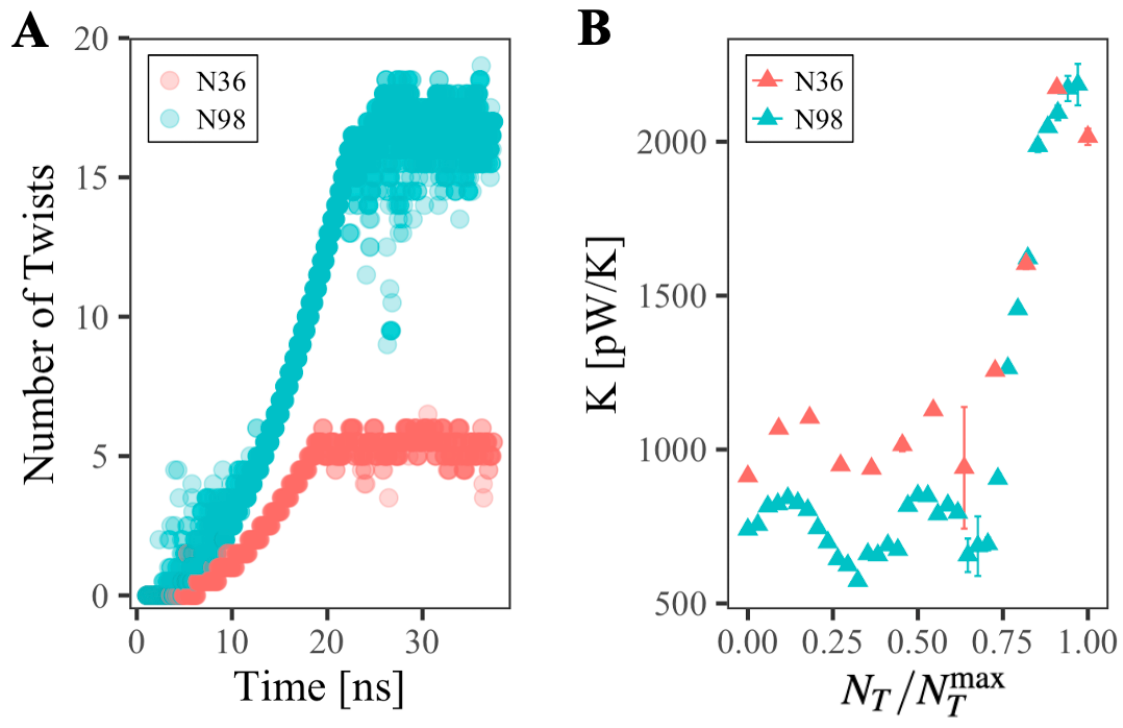}
\caption{\textbf{(A)} Twist-generating MD simulations before performing the thermal conductance measurements shown in the right panel: the number of twists in a double-helical polyethylene wire in an MD simulation where one end of the polymer is subject to a constant torque at 300K. A torque of $3\times10^{-7}~$kcal/mol was used for $N=36$ (red), and a torque of $6\times10^{-7}~$kcal/mol was used for $N=98$ (blue). Note that the number of twists for each point in this figure is computed naively based on the number of times the chains crossed, leading to some baseline fluctuations.
\textbf{(B)} The thermal conductance for the production MD simulations plotted against the twist fraction $(N_T/N_T^{\text{max}})$ for the polymer wires generated in the left panel.
The statistical uncertainties (standard deviations) are shown when larger than the symbol sizes.}
\label{FIG3}
\end{figure}

\hyperref[FIG3]{Figure \ref{FIG3}} shows that the effect of twist on the conductance behavior persists across wires of different lengths. Panel \hyperref[FIG3]{\ref{FIG3}A} shows how the number of twists increases with time when a torque is applied at one end of a polymer wire along the axis of the wire. At first $N_T$ grows quadratically as predicted by classical rotational mechanics, until a point of twist saturation $N_T^{\text{max}}$ is reached. No further increase in twist is registered beyond this point, indicating practically free rotation due to bond breakages. An important observation is that $\lambda_{\rm{twist}} = N_T^{\text{max}}(N)/N$ does not depend on $N$, implying that $\lambda_{\rm{twist}}$ is a characteristic property of a type of a polymer wire that is independent of the wire's length. When the fixed atoms at either end of the simulation box that do not contribute to the twist are accounted for, it turns out that the polyethylene double-helical wire with this force field has a value of $\lambda_{\rm{twist}} \approx 0.18$ twists per degree of polymerization. Importantly, note that the quantity $N_T^{\text{max}}$ (and hence $\lambda_{\rm{twist}}$) does not depend on the magnitude of the torque used. Also, because the torque was applied only to generate the twisted starting configurations and then turned off, the thermal conductance calculations for the twisted structures do not depend on the torque used. For comparison of different torque values and further discussion of these twist-generating simulations, see SI.

In \hyperref[FIG3]{Fig. \ref{FIG3}B} we show the heat conduction computed for configurations of different $N
(=36,98)$ but plotted against the fractional quantity $N_T/N_T^{\text{max}}.$ Clearly, the heat transport behavior observed in \hyperref[FIG2]{Fig. \ref{FIG2}} is better understood as a function of this twist fraction than as a function of the absolute number of twists. It is worth noting that the heat conductance seems to begin its spike when the twist fraction reaches $\sim0.75.$ This resembles the result of previous work which concluded that the heat conductance begins to spike when a polymer wire is stretched to a fraction $R_{\rm{frac}}\approx0.75$ of its contour length.\cite{Hadi} This suggests a curious analogy between linear and torsional strain. 

\begin{figure}[tbh]
\includegraphics[width=1\columnwidth]{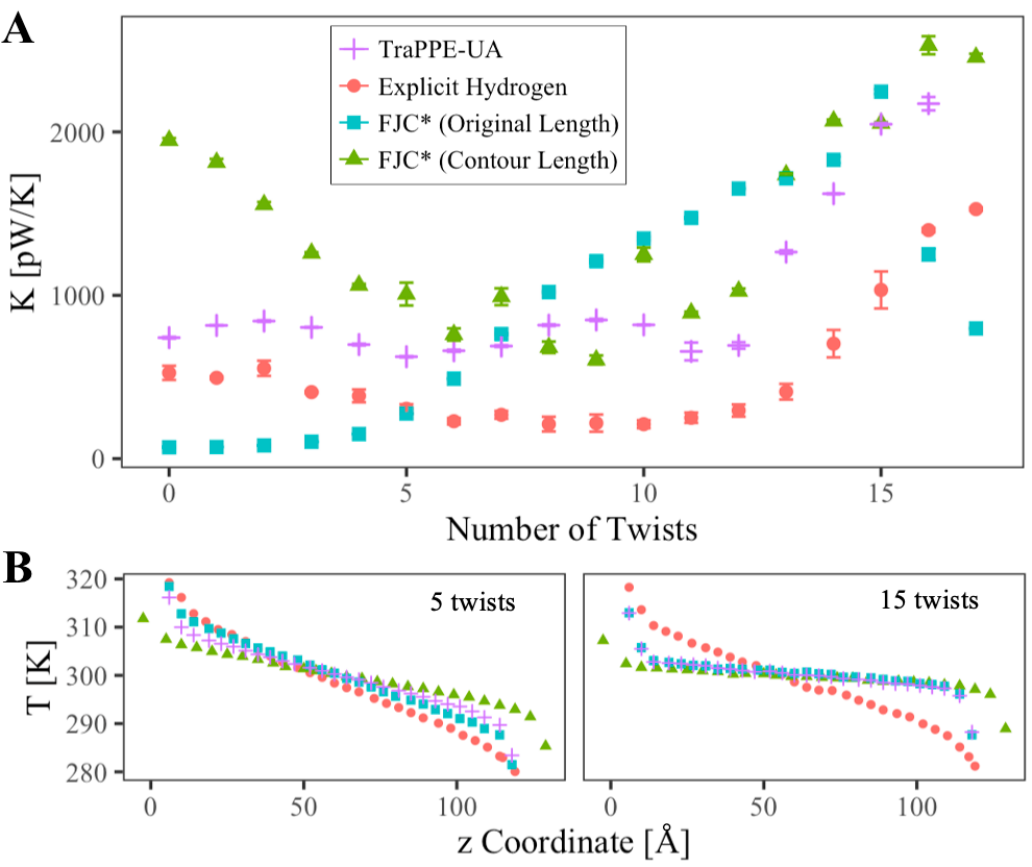}
\caption{\textbf{(A)} The thermal conductance of double-helical polyethylene wire with $N=98$ shown as functions of the number of twists for various force field models. The purple crosses correspond to the TraPPE United Atom force field and shows the same line as \hyperref[FIG2]{Fig. \ref{FIG2}} and \hyperref[FIG3]{Fig.  \ref{FIG3}}. The red circles correspond to the Explicit Hydrogen model, which treats Hydrogens as separate bodies, unlike the other models which coarse grain each CH$_x$ into a single interaction site. The blue squares correspond to the more flexible FJC model with the addition of Leonard Jones potentials to account for inter-chain repulsion. Since the FJC model omits angular potentials, the CCC bond angles can relax to $180^{\circ}$. As such, the natural length is longer than in our original force field; this adjusted length is used in the model corresponding to the green triangles.~\textbf{(B)} Average temperature profiles along the direction of the wire (z coordinate) for selected levels of twist, where the colors and shapes have the same meaning as panel A.}
\label{FIG4}
\end{figure}

\hyperref[FIG4]{Figure \ref{FIG4}} shows similar results for several different force field models for length $N=98$. The TraPPE-UA FF used in the previous
figures, which coarse grains each CH$_x$ monomer into a single interaction site, is indicated by purple crosses. The Explicit Hydrogen model, which does not make this united atom simplification, is indicated by red circles. The other two models refer to the FJC$^*$ model which omits the angular and dihedral potentials. For the simulations indicated by the blue squares, the end-to-end distance is the same as the first two models. For the green triangles, the end-to-end distance of the polymer has been increased to account for the bond angle relaxation allowed by the FJC$^*$ model as discussed in \hyperref[details]{Sec. II}. The following observations are noteworthy:

\begin{quote}
(a) The results for the TraPPE-UA and Explicit Hydrogen models are quite similar for 0 twists. Differences increase with twist, suggesting that modeling twisted or coiled structures could be a relative weakness of UA models. This could be due to the detailed steric interactions of the hydrogen atoms between chains that are not accounted for when the UA simplification is made. Note that the TraPPE-UA and Explicit Hydrogen models used in this paper are developed independently from each other; previous works such as Ref. \citen{unitedWorks} comparing the UA FF with the same all-atom FF results have found agreements in the thermal conductances within the statistical uncertainties for single, untwisted molecules. Our purpose here is not to assess the accuracy of UA models but to assess qualitatively whether the trends observed persist with these models.\\
(b) Despite the deviations with increased twist, the main qualitative observations from our discussion of \hyperref[FIG2]{Fig.  \ref{FIG2}} persist in both the TraPPE-UA and Explicit Hydrogen models. These are the presence of oscillations as we increase twist, and a large spike as we approach twist saturation. \\
(c) The FJC$^*$ models strongly deviates from the other two. This suggests that angular and dihedral contributions to the force field are important in twist simulations. Nevertheless, in these models as well, the heat conductance shoots up at high levels of twist.\\
(d) For untwisted wires, the thermal conductance in the FJC$^*$ model without making the length adjustment (blue squares) is quite low. This is consistent with our previous finding that slack greatly diminishes the heat conductance in polymer wires compared to their natural length.\cite{Hadi}\\
(e) Here we have used 300 K as the average temperature of the polymer wire, but we find that the trends observed persist when the same calculation is performed with different average temperatures (results shown in SI).

\end{quote}

Overall, we see that the high-twist increase is present in all models, but the pattern of oscillation in the low-twist regime varies.
This raises the possibility that two competing factors affect the thermal conductance as a polymer wire is twisted: (i) increase in torsional strain which could increase the thermal conductance in all force fields, and (ii) damping of normal modes important for ballistic transport caused by the interlocking of the two chains, which could decrease the thermal conductance. The latter could be highly FF dependent, and the oscillations in the low-twist regime could be attributable to subtleties of the competition between these two factors. Some evidence for this is seen in the delocalization-weighted density of modes plots discussed below.

{\it Thicker wires.} 

\begin{figure}[tbh]
\includegraphics[width=1\columnwidth]{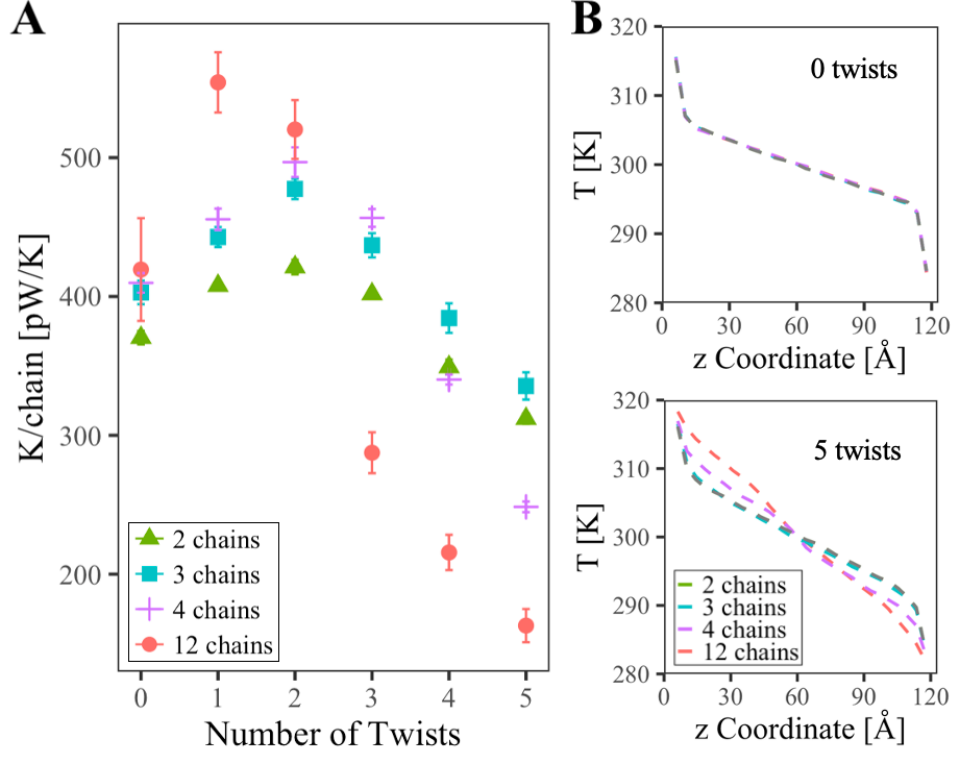}
\caption{\textbf{(A)} The thermal conductance per chain plotted as a function of the number of twists for wires consisting of 2 (green triangles), 3 (blue squares), 4 (purple crosses), and 12 (red circles) polyethylene chains of $N=98$. Results are shown for up to 5 twists; note that green triangles are points plotted in \hyperref[FIG2]{Fig. \ref{FIG2}} and \hyperref[FIG2]{Fig. \ref{FIG3}} (except scaled by a factor of 0.5 here). \textbf{(B)} The average temperature profiles along the direction of the wire ($z$ coordinate) for the same wires containing 0 twists (top) and 5 twists (bottom).}
\label{FIG5}
\end{figure}

The fact that thermal conductance in polymer wires seems to be sensitive to twist raises the possibility that twist can be used to engineer polymer wires with improved thermal conduction. Therefore, it is worth exploring the extent to which the thermal conductance behavior found in the double helix is generalizable to wires of arbitrary thickness. Using the TraPPE-UA force field, \hyperref[FIG5]{Fig. \ref{FIG5}} shows the per-chain conductance for twisted $N=98$ polyethylene wires composed of 2 (green triangles), 3 (blue squares), 4 (purpled crosses), and 12 (red circles) chains, as well as selected temperature profiles. It should be noted that the issue of twist saturation described above is more complex in multichain ($>$2) wires as partial bond-breaking starts to occur before a total softening of the structure under twist is observed (see
SI). For this reason, we report results here for only up to 5 twists, corresponding to the first oscillation in \hyperref[FIG2]{Fig. \ref{FIG2}.} 

The main observation of \hyperref[FIG5]{Fig. \ref{FIG5}A} is that the first oscillation observed for two chains does persist for chains of various thicknesses, and the oscillation actually seems to increase in amplitude for the thicker wires. Remarkably, when just one twist is applied to the 12-chain wire, the conductance increases by over 30$\%$. Since applying slightly more twist actually decreases the thermal conductance below its original value, this raises the possibility that precisely tuning the amount of twist in a wire could be consequential.

{\it Correlations with Other Observables.}

\begin{figure}[tbh]
\includegraphics[width=1\columnwidth]{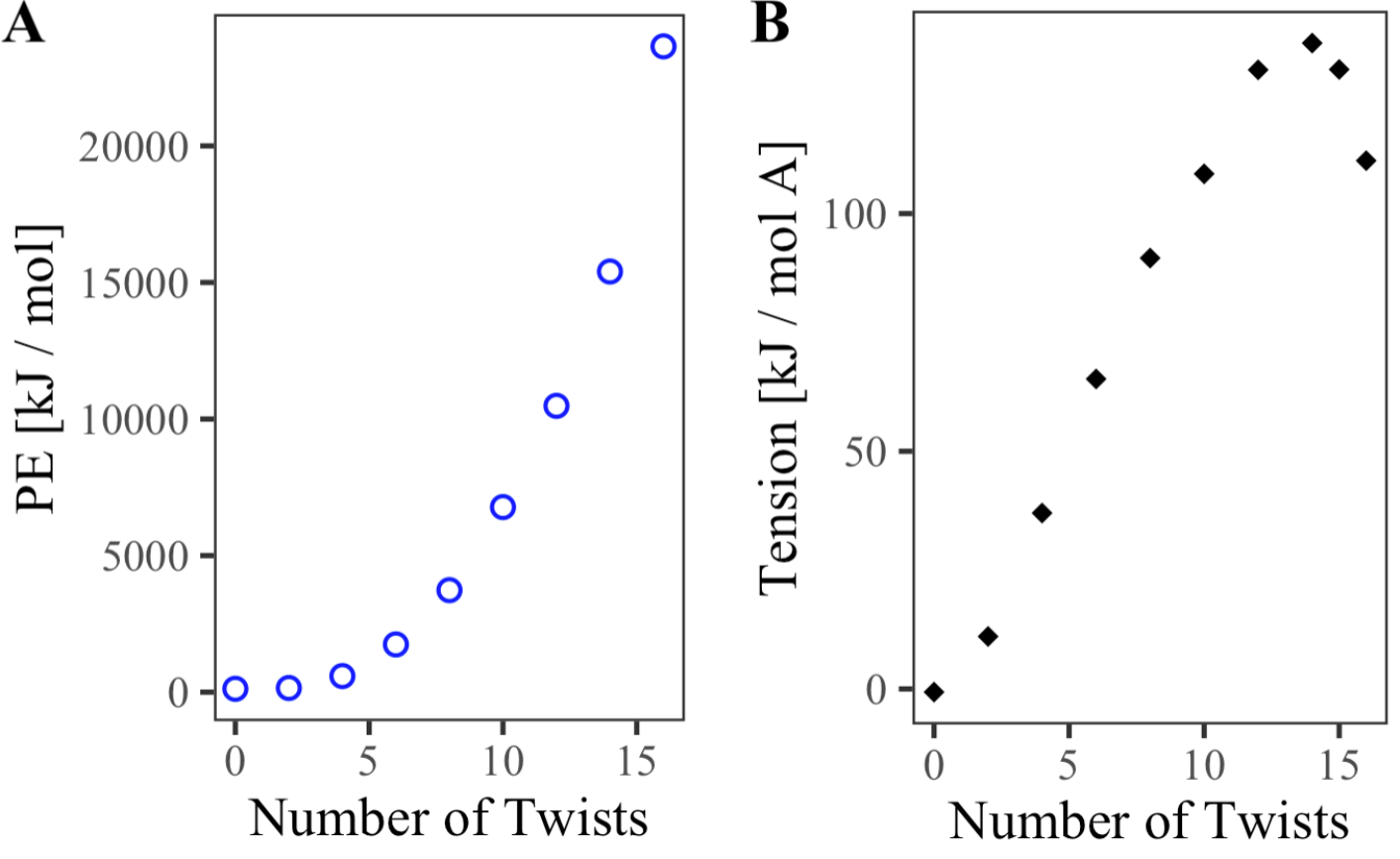}
\caption{\textbf{(A)} The potential energy in a double-helical polyethylene wire of $N=98$ plotted as a function of the number of twists. \textbf{(B)} The linear tension, calculated as the derivative of the potential energy upon infinitesimal stretching, in the same polymer wire.}
\label{FIG6}
\end{figure}

In order to further understand the structural effects of twist on polymer wires, we looked at the effect of twist on several other physical observables using the two-stranded wire. \hyperref[FIG6]{Figure \ref{FIG6}A} shows the total potential energy in the polymer wire as a function of twist. The quadratic nature of the PE-twist relationship suggests that the wire stores energy in a spring-like manner as it is twisted. \hyperref[FIG6]{Figure \ref{FIG6}B} shows the linear tension in the wire, obtained from $d(PE)/dL$ where $dL$ is a small stretch ($L \rightarrow L + dL)$ from the natural length $L.$  In \hyperref[FIG6]{Fig. \ref{FIG6}B}, the most notable feature is that while the tension increases roughly linearly in the low-twist regime, there comes a point in the high-twist regime when increasing the twist actually reduces the tension. It is interesting to note that this transition from increasing tension to decreasing tension, and the steep increase in potential energy, both occur in the twist regime where the thermal conductance spikes (see \hyperref[FIG2]{Fig. \ref{FIG2}}). Also note that this counterintuitive effect of increased twist decreasing the tension is reminiscent of an observation by Lionnet \textit{et al.}\cite{DNAtwist} that overtwisting a DNA strand causes it to lengthen.

\begin{figure}[tbh]
\includegraphics[width=1.01\linewidth]{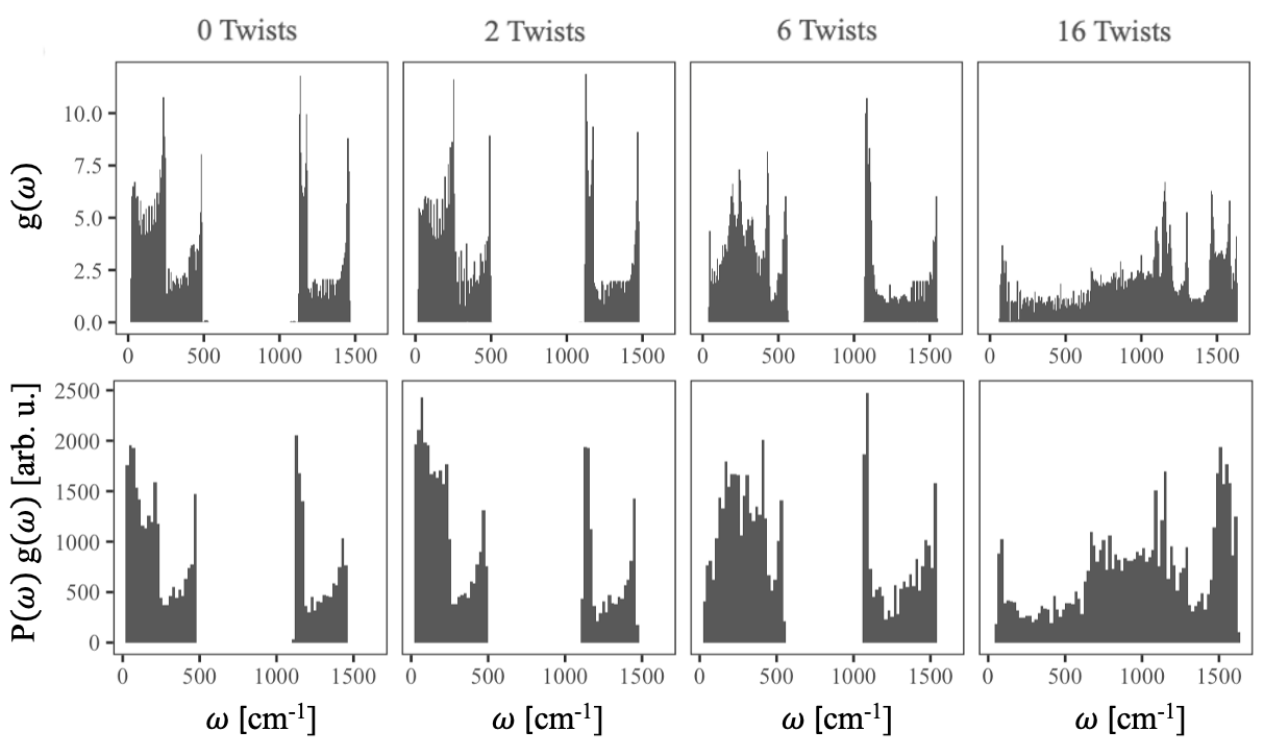}
\caption{Spectral densities $g(\omega)$ defined as the density of modes (top), and the product $P(\omega)g(\omega)$ with $P(\omega)$ being the participation ratio (bottom) plotted against the frequency $\omega$.  The structures are $N=98$ double-helical polyethylene wires using the TraPPE-UA force field and containing various numbers of twists. The quantity $g(\omega)$ is the number of modes in each bin  (unnormalized). $P(\omega)$ is $\Sigma_k P_k$ in the bin divided by the number of modes in the bin.}
\label{FIG7}
\end{figure}

In an attempt to gain more insight into the origin of the heat conductance behaviors described above, we have also examined the spectral densities and localization properties of the normal modes for some structures. At room temperature, phonons of low frequency are thought to play the greatest role in heat transport both because they are thermally populated and because they are less subjected to local scattering events. The top row of \hyperref[FIG7]{Fig. \ref{FIG7}} shows the spectral density $g(\omega)$ of normal modes as a function of frequency $\omega$ for selected amounts of twist. Note the gap in the mid-frequency range of the spectral density that disappears at high twist. As we discuss in the SI, the disappearance of this gap is related to the intermolecular interactions between the chains: the top panel of Fig. S6 shows that when the well depth of the LJ interaction is turned down for $10$ twists, gaps begin to appear in the spectral density.
In the bottom row of \hyperref[FIG7]{Fig. \ref{FIG7}}, we scale the spectral density by the average participation ratio in each bin (see Eq. \ref{eq:Pk}). We see that the product $P(\omega) g(\omega)$ does seem to predict the oscillatory pattern of heat conductance in the low twist regime. In the lowest frequency part of the spectrum ($\omega < 250$ cm$^{-1}$), the area under the curve grows between 0 twists and 2 twists as the thermal conductance rises and then falls at 6 twists where the thermal conductance is lower (compare with \hyperref[FIG2]{Fig. \ref{FIG2}}). 
However, this analysis of the spectrum above does not seem to predict the high thermal conductance obtained in the most twisted structures. Specifically, the thermal conductance is by far the highest for 16 twists, yet the product $P(\omega) g(\omega)$ is by far the lowest (again referring to $\omega < 250$ cm$^{-1}$). Further evidence that $P(\omega) g(\omega)$ does not always correlate with thermal conductance is presented in Table S1 and Fig. S5 in the SI, which show that when interchain interactions are turned down the low frequency $P(\omega) g(\omega)$ increases but thermal conductance decreases. Overall, we find that while analysis of the the $P(\omega) g(\omega)$ can explain the low-twist oscillations, the smoothness of the helical curves (see \hyperref[FIG8]{Fig. \ref{FIG8}}) is a better predictor of the high twist spike.

\begin{figure}[tbh]
\includegraphics[width=1\columnwidth]{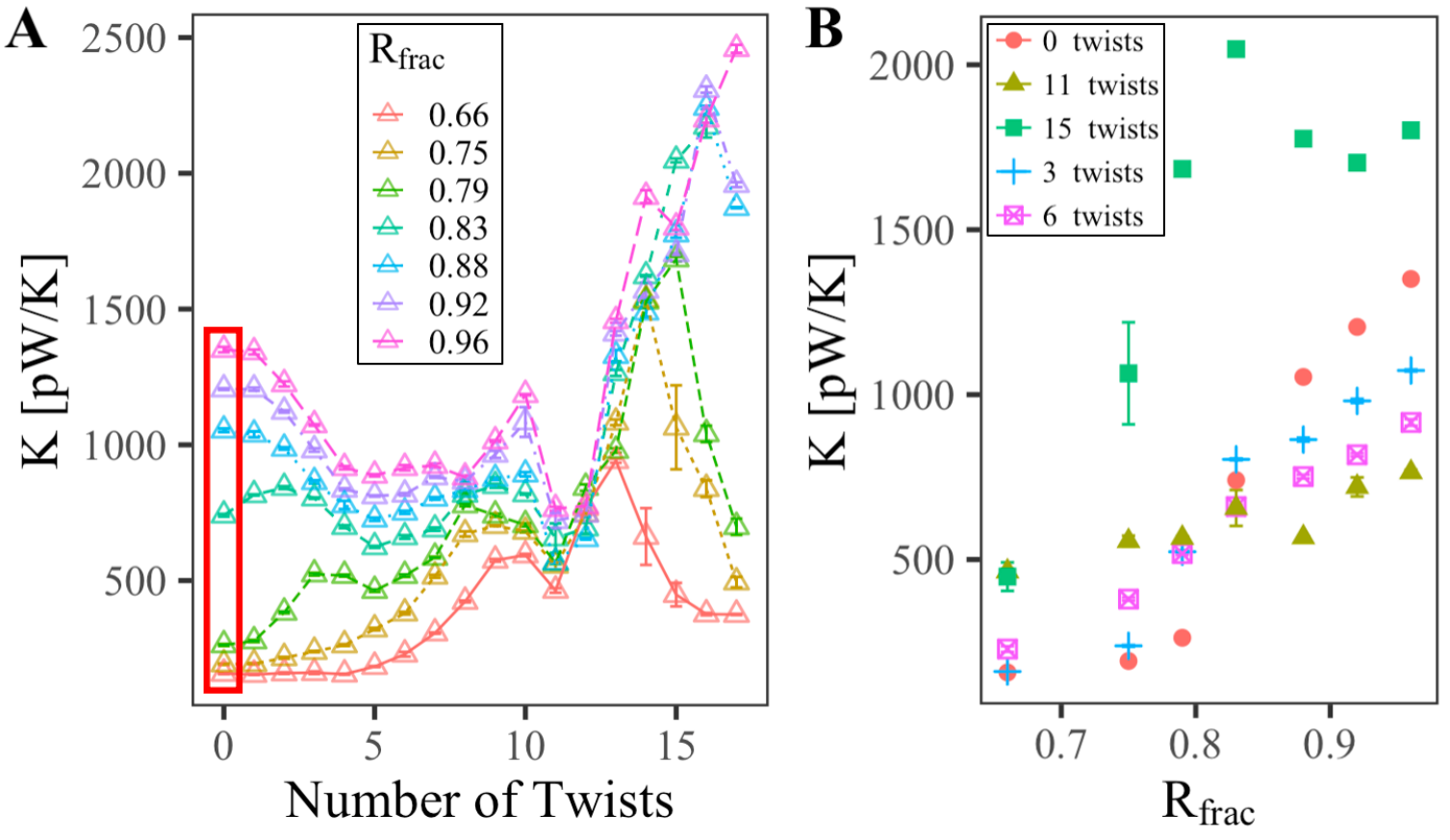}
\caption{\textbf{(A)} The thermal conductance plotted as a function of twist in a double-helical polyethylene wire with $N=98$ for selected values of R$_{\text{frac}}$ = R$_{\text{ete}}$/R$_{\text{contour}}$. \textbf{(B)} The thermal conductance of the same wires as the left panel, plotted as a function of R$_{\text{frac}}$ = R$_{\text{ete}}$/R$_{\text{contour}}$ for selected levels of twist. The points in the red box in (A) are the red circles in (B).}
\label{FIG9}
\end{figure}

{\it Combining stretch and twist.} Previous work has shown an increase in thermal conductance when a polymer wire is stretched,\cite{Hadi} while here we have found interesting behavior when the same wire is twisted. \hyperref[FIG9]{Figure \ref{FIG9}} shows the result of combining stretch and twist in our model $N=98$ double helix. The left panel shows the thermal conductance as a function of twist for various stretch levels $R_{\rm{frac}}=R_{\rm{ete}}/R_{\rm{contour}},$ where $R_{\rm{contour}}$ is the contour length of the polymer and $R_{\rm{ete}}$ is the given end-to-end distance. The right panel shows the dependence of thermal conductance on the $R_{\rm{frac}}$ for various levels of twist. The two plots can be viewed as transforms of each other with respect to the same data. For instance, the red circles in the right panel are obtained from the column of points highlighted by the red box (see above 0 twists) on the left panel. Note that the natural length of the polymers corresponds to $R_{\rm{frac}} = 0.83.$

Some observations include the fact that the high-twist spike is present for all levels of stretch, though for the shorter $R_{\rm{frac}}$ values, kinks and bond breakages seem to appear at high levels of twist causing the observed drop-offs. Also the relative minimum at 11 twists persists for all levels of stretch. In the low-twist regime, the oscillations are different but seem to be correlated\footnote{The manner in which these curves seem to be related could be described by a path homotopy. In topology, a path homotopy is a function that continuously deforms a path $f$ into a path $g$ according to a parameter. Here that parameter would be $R_{\rm{frac}}$} in a non-trivial manner.
\cite{munkrestopology} Finally, we note that when comparing a highly twisted structure with a moderate amount of stretch to an untwisted structure with a moderate amount of slack, the thermal conductance can differ by well over an order of magnitude.

{\it Chirality Considerations.}

Inspired by the recent interest in chirality effects on molecular transport properties, we also applied a mathematical tool for quantifying chirality, the Continuous Chirality Measure (CCM), in search for possible correlations between thermal conductance and chirality. The CCM computes an abstract distance of a structure from its \textit{nearest achiral structure}, obtained by averaging the original structure with its enantiomer. It ranges from 0 for an achiral structure to 1 for a structure with high chirality. As we will discuss further in future work, it makes most sense to apply the CCM locally, on a length scale $\tau$, so that chirality depends on the intrinsic geometry of the helix rather than the chain's length. This approach is consistent with the mathematical work of Raos,\cite{helicalRibbons} who has explored how the CCM of helical ribbons is affected by geometric parameters (radius $a$ and pitch $h$).

\begin{figure}
\includegraphics[width=1\columnwidth]{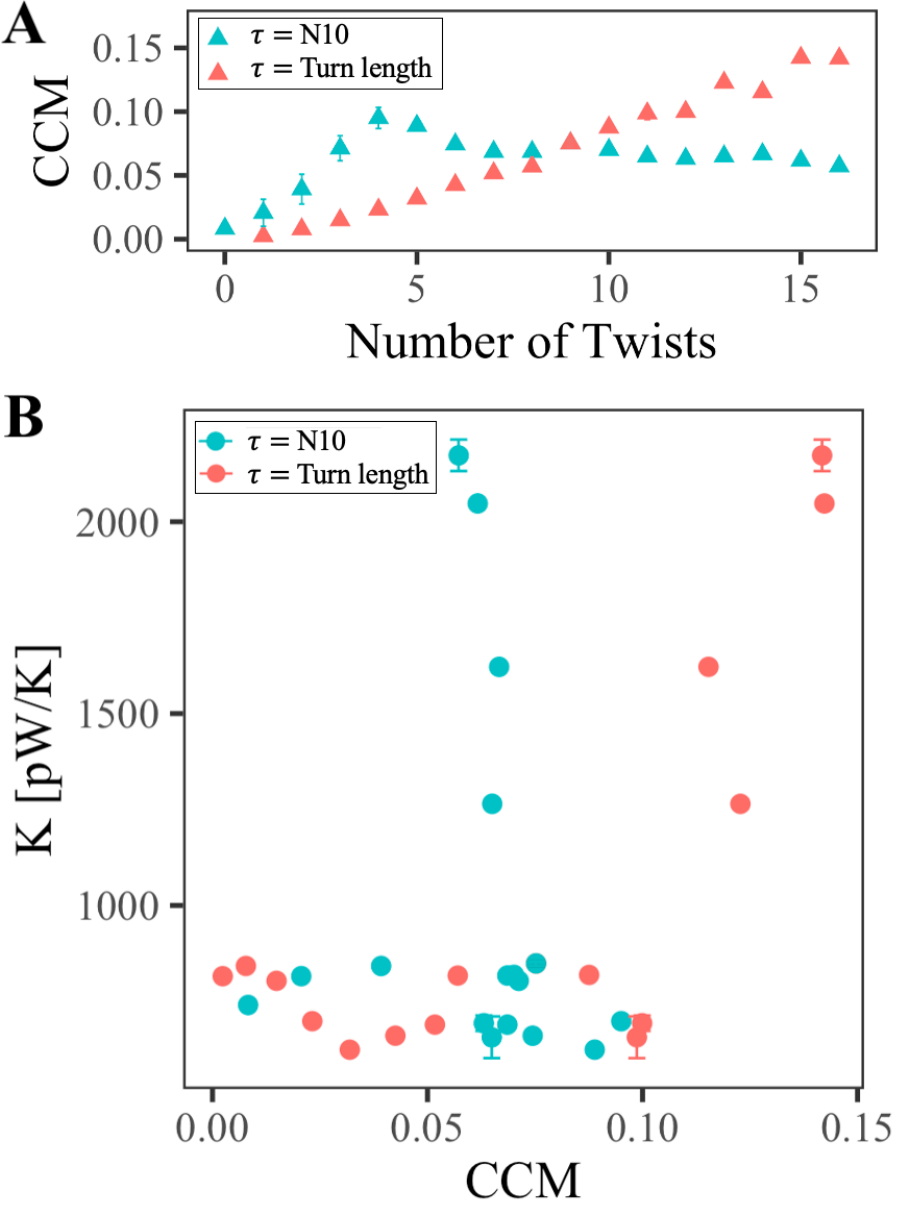}
\caption{\textbf{(A)} Continuous Chirality Measure (CCM) plotted as a function of the number of twists in a double-helical polyethylene wire of $N=98$ when the molecular structure is considered to be a segment of the wire 10 atoms long (blue), or one helical turn (red). See SI for mathematical details of the CCM. \textbf{(B)} Thermal conductance as a function of CCM for the same two length scales as above. The statistical uncertainties (standard deviations) are shown when larger than the symbol sizes.}
\label{FIG10}
\end{figure}

\hyperref[FIG10]{Figure \ref{FIG10}A} shows how the CCM is affected by the number of twists for two such choices $\tau.$ The first choice is a fixed length corresponding to 10 atoms along the chain (blue). The second choice for $\tau$ is the length of a complete helical turn (red), which varies based on the wire's amount of twist. We see that with the latter approach, the CCM grows monotonically with the number of twists, while with the former approach, the CCM obtains a maximum and then decreases slightly for greater amounts of twist. \hyperref[FIG10]{Figure \ref{FIG10}B} plots thermal conductance as a function of the CCM using both approaches. Note that since the red line ($\tau$ = \text{Turn length}) in \hyperref[FIG10]{Fig. \ref{FIG10}A} appears linear, the red circles in \hyperref[FIG10]{Fig. 10B} display a similar pattern as when the horizontal axis is the number of twists (compare to \hyperref[FIG2]{Fig. 2}). The other approach ($\tau=N10$) does not show any clear pattern. While our results clearly indicate (as expected) that both chirality (as measured by the CCM) and heat conduction depend on the level of twisting, our simulations do not show a clear correlation between these two observables. While it is true that heat conductance depends on twist and chirality depends on twist, this does not implicate chirality \textit{per se} as a causal agent in the former. Claims have been made that chirality affects thermal conductance in Carbon nanotubes.\cite{Nanotubes1,Nanotubes2,Nanotubes3,Nanotubes4,Nanotubes5, Nanotubes6, Nanotubes7} However, these claims used factors such as structural angles or optical activity as proxies for chirality. These factors may correlate with both chirality and observed transport behaviors, but this does not imply causation. It is likely that the observed increase in thermal conductance was caused by structural factors other than chirality (see \hyperref[FIG8]{Fig. \ref{FIG8}}).

\section{Summary \& Conclusions}
\label{conclusion}

We have used non-equilibrium MD simulations to study nanoscale control of thermal conductance along polymer wires, and we have found a high sensitivity to twist. For the two-stranded polyethylene wire, we found oscillations in the thermal conductance at relatively low levels of twist, and a substantial rise in thermal conductance at high levels of twist just before the breakage of the wire occurs. These trends appeared in all force fields and levels of stretch examined, though the details varied. The fact that the high-twist spike persists across various force fields
and levels of stretch while the subtleties of the oscillations varied suggests
two separate mechanisms by which twist affects the thermal conductance, one that is more general and one of which is more subtle and sensitive to details of the particular polymer. We examined the normal mode spectrum and found that when mode delocalization was considered, the oscillations in the low-twist regime could be predicted. However, the normal mode spectrum in the high-twist regime did not always correlate with the observed increase in thermal conductance.
Nevertheless, the high-twist spike turns out to be correlated to the structural transition from zig-zag patterns to smooth helical curves, suggesting that twist can alter the ordering of atoms in the polyethylene polymer wires, affecting the phase of the propagating phonons and therefore the scattering center details. We found that the threshold for the transition from the low-twist oscillations to the high-twist spike appeared to be a fraction $\sim 0.75$ of twist saturation.  Interestingly, this was also the threshold fraction of contour length at which heat conductance began to spike when end-to-end distance was controlled in a previous study. \cite{Hadi} Finally, we have used the CCM, a mathematical tool that treats chirality as a continuous rather than binary property, to examine the chirality of the double-helical wire. Though we did not find correlations between thermal conductance and chirality independent from that with the number of twists, we demonstrate the possibility of using quantitative methods to study molecular chirality, which has previously been a rather qualitative and hand-wavy subject. Overall this study strongly suggests that applying twists to polymer wires could be a useful strategy for engineering wires with high thermal conductance. 

\section*{SUPPLEMENTARY INFORMATION}
The details about twist-generating simulations, effects of intermolecular interactions, temperature, and number of polymer chains on the thermal conductance, and details of the explicit hydrogen model used are presented in the Supplementary Information.

\section*{Acknowledgements}
The research of A.N. is supported This material is based upon work supported by the Air Force Office of Scientific Research under award number FA9550-23-1-0368 and the University of Pennsylvania. E.A. aknowledges the support of the the University of Pennsylvania (GfFMUR Grant). M.D. acknowledges the support from by U.S. Department of Energy, Office of Science, Basic Energy Sciences, Chemical Sciences, Geosciences, and Biosciences Division, Condensed Phase and Interfacial Molecular Science program, FWP 16249. Pacific Northwest National Laboratory (PNNL) is operated by Battelle for the U.S. DOE under Contract No. DE- AC05-76RL01830. Computing resources were partly allocated by PNNL\textquoteright s Institutional Computing program. C.C. was supported by a postdoctoral fellowship from the Vagelos Institute for Energy Science and Technology (VIEST). The authors are grateful to Zeyu Zhou and Abel Carreras for technical assistance and discussions and also to Pere Alemany, David Avnir and Inbal Tuvi-Arad for discussions on continuous chirality measures. 

\section*{Data Availability}
The data that support the findings of this study are available on github at: \url{https://github.com/eabes23/polymer_twist/}. 

\newpage    
\pagebreak  
\newpage

\clearpage
\section*{References}
\bibliographystyle{aipnum4-1}
%

\pagebreak
\widetext
\begin{center}
\textbf{\large Supplemental Information: Heat Transport with a Twist}
\end{center}
\setcounter{equation}{0}
\setcounter{figure}{0}
\setcounter{table}{0}
\setcounter{section}{0}
\setcounter{page}{1}
\makeatletter
\renewcommand{\theequation}{S\arabic{equation}}
\renewcommand{\thefigure}{S\arabic{figure}}
\renewcommand{\bibnumfmt}[1]{[S#1]}
\renewcommand{\citenumfont}[1]{S#1}
\renewcommand{\thepage}{S\arabic{page}} 

\section{Proof of Principle for Twist-Generating Simulations}
\label{appendix:genTwist}

The twisted structures used in this study were obtained from the structures of untwisted polymer wires by performing twist-generating simulations in which the atoms on one end of the wire were immobilized while the atoms on the other end of the wire were treated as a rigid body subject to a constant torque. We found that the twist at first increased quadratically as predicted by classical rotational dynamics until a characteristic point of twist-per-length saturation was reached. The coordinates at various snapshots that had the desired number of twists could then be used as input files for our thermal conductance simulations. Since this twist-generating simulation is both novel and imperative to the validity of our study, we have included this appendix for further justification. 

\begin{figure*}[h]
\includegraphics[width=0.9\linewidth]{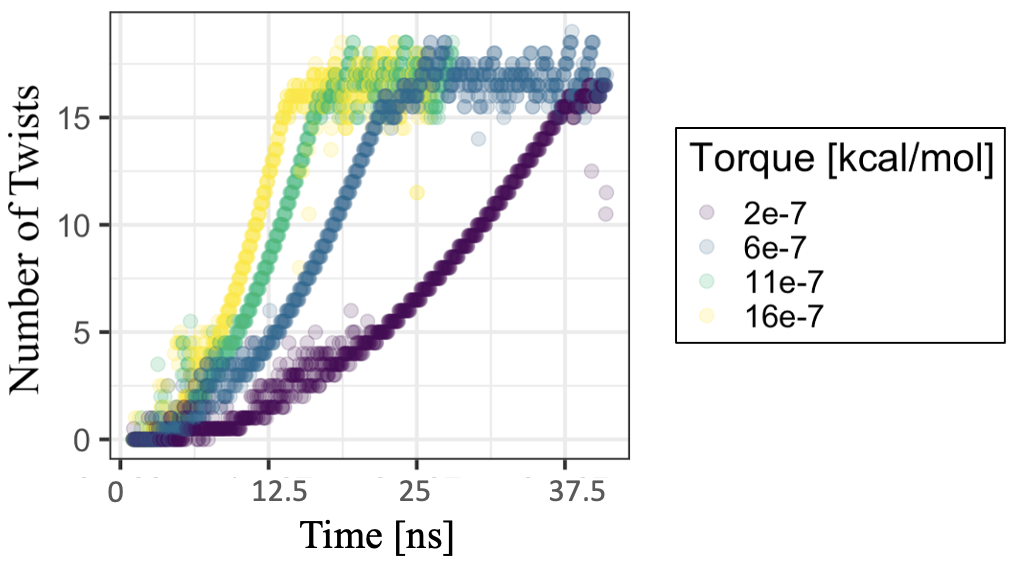}
\caption{The number of twists in a double-helical polyethylene $N=98$ wire in an MD simulation where one end of the polymer is subject to a constant torque at 300K. Results are shown when the applied torque is equal to $2 \times 10^{-7}$ kcal/mol (purple), $6 \times 10^{-7}$ kcal/mol (blue), $11 \times 10^{-7}$ kcal/mol (green), and $16 \times 10^{-7}$ kcal/mol (yellow).}
\label{FIG11}
\end{figure*}

\hyperref[FIG11]{Figure \ref{FIG11}} shows that the same twist saturation is reached for several amounts of applied torque. It should be noted that this is the case only because all of the torques shown used for this figure are very strong relative to the polymer wire and easily strong enough to break bonds. As implied by the spring-like relationship between potential energy and twist shown in Fig. 7 of the main text, there is a torque regime (perhaps orders of magnitude smaller than those used in these simulations) where different equilibrium amounts of twist would be reached. However, using a large torque to quickly twist-saturate the wire was fast and allowed us to obtain structures of all levels of twist from a single simulation.

\begin{figure*}[h]
\includegraphics[width=0.9\linewidth]{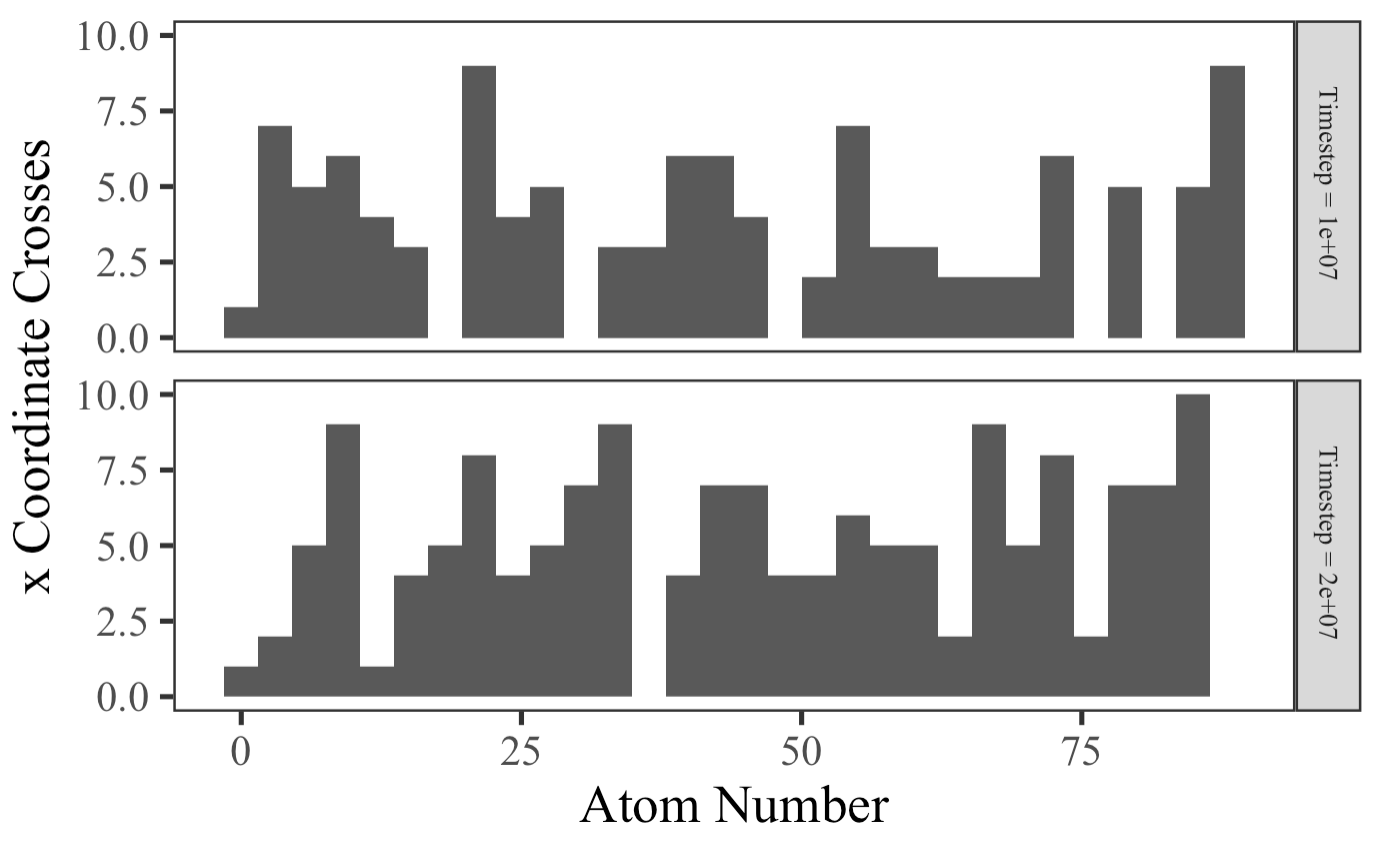}
\caption{Histogram of the number of times the two strands cross in the x-direction (where the $z$ direction is the length of the polymer) binned by the monomer number where the cross occurs in twist-generating simulations for a double-helical polyethylene wire with torque of $6 \times 10^{-7}$ at 300K. The bin totals are aggregated from several such simulations in order to generate enough data for the bins to be reasonably populated. Results are shown for two distinct timesteps throughout the simulation, where the timestep length is 1.25 fs. The timesteps are $1\times10^7$ (top panel) and $2\times10^7$ (bottom panel).} 
\label{FIG12}
\end{figure*}

A second concern might be the asymmetry of the twist-generating simulation: since torque is applied to only one end, the concern might be raised that the result will be an asymmetric structure, with a higher density of twists and/or higher particle kinetic energies on the side where the torque is applied. \hyperref[FIG11]{Figure 11} and \hyperref[FIG12]{Fig. \ref{FIG12}} support that this is not the case for the TraPPE-UA model. \hyperref[FIG11]{Figure \ref{FIG11}} shows that the twists are evenly distributed throughout the chain at multiple MD steps throughout the simulation. The x-coordinate-crosses refer to the number of times the difference in the x-coordinate values between corresponding atoms on the two chains changes sign when stepping along the length of the chain. The number of twists is then approximated as half the number of x-coordinate-crosses. The atom number refers to the identity of the atoms (between 0 and 98 for $N=98$ length chains) where such a cross occurs. The data in \hyperref[FIG11]{Fig. \ref{FIG11}} is binned over several independent twist-generating simulations to aggregate enough data to make the histogram meaningful.

\begin{figure*}[h]
\includegraphics[width=0.9\linewidth]{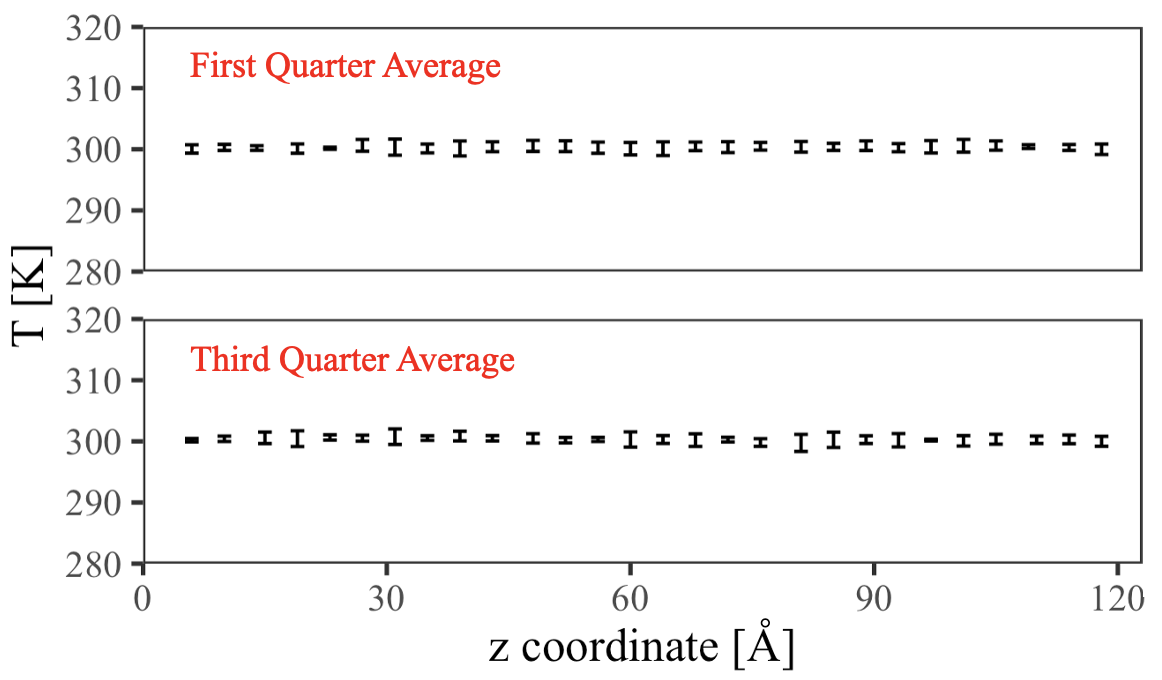}
\caption{The average temperature profiles obtained for the first and third quarters of the simulation described in a twist-generating simulation for a double-helical polyethylene wire with a torque of $6 \times 10^{-7}$ kcal/mol at 300K. The total simulation length is $32\times10^7$ timesteps of 1.25 fs each.}
\label{FIG13}
\end{figure*}

\hyperref[FIG13]{Figure \ref{FIG13}} shows average temperature profiles of the first and third quarters of a twist-generating simulation, obtained in the same way as in the main figures of this paper. This figure indicates that the asymmetry of the applied torque does not lead to an uneven energy distribution throughout the chain or any departure from thermal equilibrium. 

\clearpage

\section{Effect of Intermolecular Interactions on the Thermal Conductance}

In this Section, we study the significance of intermolecular interactions between the two chains in the thermal conductance behavior of double-helical polymer wires
as a way to get more insight into the relationship between structure and heat conduction in the studied system.
We note that the structures of the twisted polymer chains are determined by a competition between intermolecular and intramolecular interactions.
We find (see details below) that a main factor in the variation of heat conduction is a transition from a zig-zag sub-structure charecteristic of a typical polyethelene chain to a smoother helical structure. The zig-zag substructure, as compared to the smoother helical structure, would have increase phononic backscattering and consequently reduced heat conduction.

\begin{figure*}[tbh]
\includegraphics[width=\linewidth]{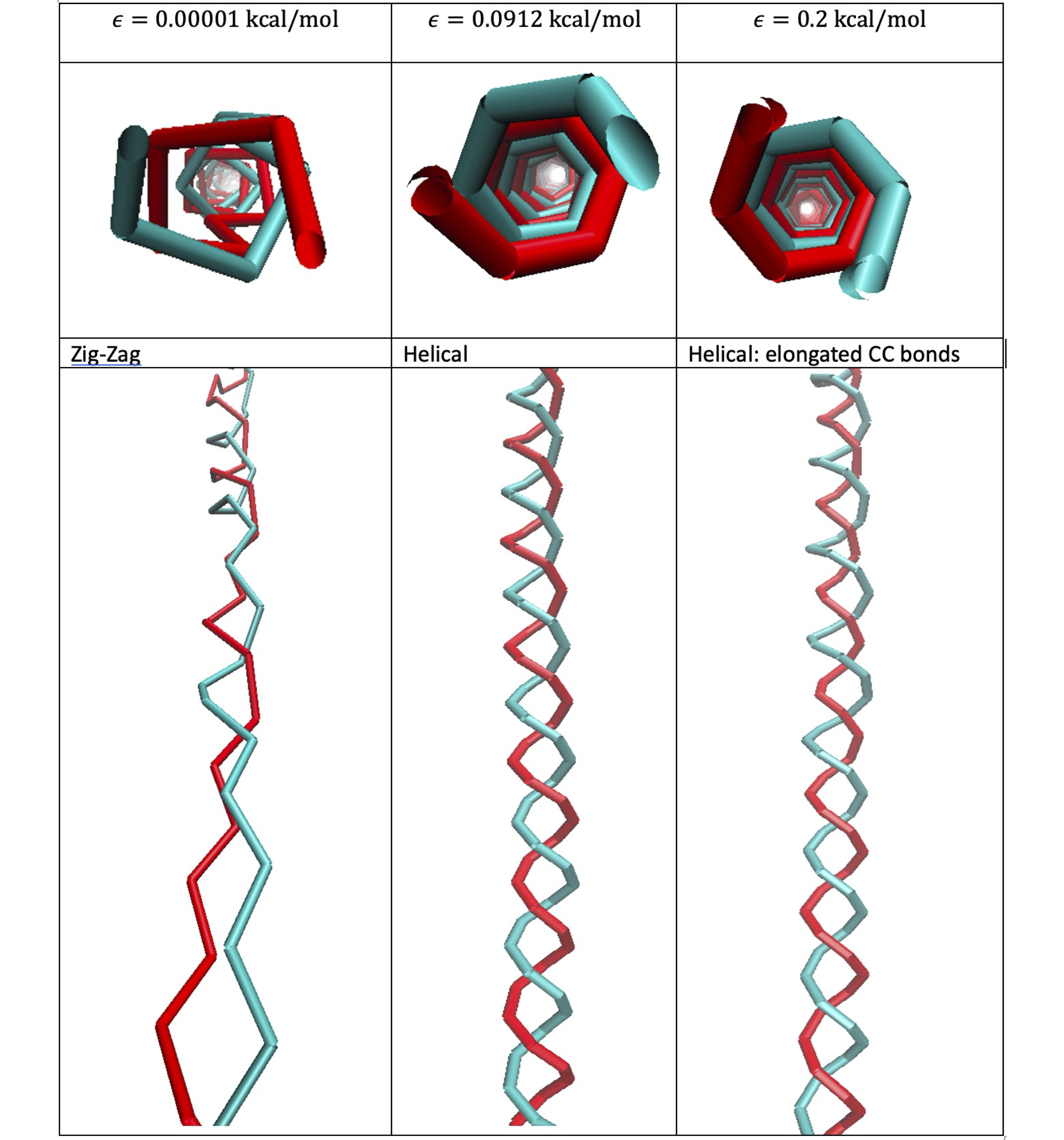}
\caption{The snapshots of two polymer chains, each consisting of $98$ units at $300$ K when the number of twists is $16$ when the intermolecular interactions (LJ potential parameters) change. Note that $\sigma$ for all intermolecular interactions is $3.95$ \AA . When the well depth magnitudes are very small ($\epsilon=0.00001$ kcal/mol) the polymer chain conformations are more random resembling zig-zag structures. More uniform helical structures are observed for well depth magnitudes of $0.0912$ and $0.2$ kcal/mol. The average CC bond length increases when the well depth magnitude is $0.2$ kcal/mol as compared to the one from the well depth magnitude of $0.0912$ kcal/mol. Note that one polymer is shown in red and the other one is shown in cyan. }
\label{snap-inter}
\end{figure*}

We focus on polymer chains of length $N=98$ with selected number of twists ($10$ and $16$ twists). Table \ref{inter} shows the thermal conductance values for two polymer chains with $16$ number of twists as the well depth magnitudes change. Fig. \ref{snap-inter} also shows the snapshots of the conformations of the polymer chains at three selected well depth magnitudes.

\begin{table}[tbh]
\centering
\caption{Thermal conductance values, $K$ , for $2$ polymer chains with $16$ number of twists as the intermolecular interactions change. The intermolecular potential is the LJ potential with $\epsilon$ as the well depth and $\sigma$ as the parameter for the hard sphere diameter. For all calculations here, $\sigma = 3.95$ \AA\ while the $\epsilon$ values change.}
\begin{tabular}{|c|c|c|}
     \hline
     $\epsilon$, kcal mol$^{-1}$ &  $K$, pW/K  \\
     \hline
     $10^{-5}$ & $296\pm5$  \\
     $10^{-3}$ & $798\pm8$   \\
     $0.0912$  & $2315\pm20$  \\
     $0.2$     & $876\pm6$  \\
     $0.6$     & $832\pm10$  \\
     $0.9$     & $762\pm12$  \\
     \hline
\end{tabular}
\label{inter}
\end{table}

As can be seen, the dependence on $\epsilon$ is not monotonic, and $K$ decreases when $\epsilon$ becomes considerably smaller or larger than the value $\epsilon = 0.0912$ kcal/mol  which is the TraPPE well depth magnitude. A visual inspection of the molecular structure (e.g., a snapshot shown on the left side of Fig. \ref{snap-inter}) shows that in the small $\epsilon$ regime the molecular double-helical chain develops a zig-zag substructure that may cause more phonon (back)scattering events and consequently smaller thermal conductance. In contrast, in the large $\epsilon$ regime, tight intermolecular interactions results in increased average CC bond lengths disrupting the optimum helical structure for enhanced thermal conductance (snapshot shown on the right side of Fig. \ref{snap-inter}).

\clearpage

We have also calculated the spectral densities of modes of twisted pair polymer chains with $16$ number of twists when the intermolecular interactions change from very small well depth magnitudes to relatively large well depth magnitudes. 

\begin{figure*}[tbh]
\includegraphics[width=0.9\linewidth]{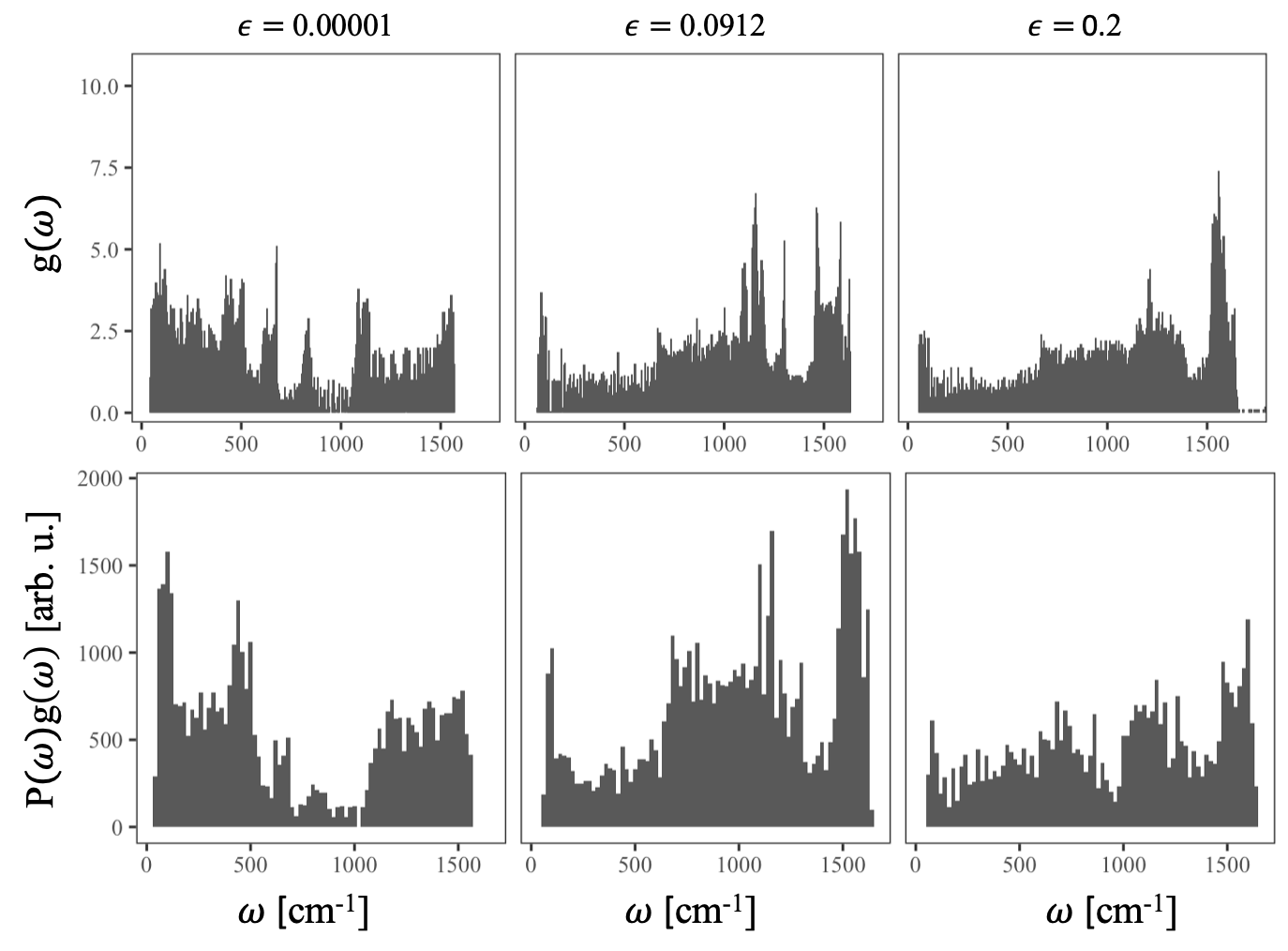}
\caption{The spectral densities of modes, $g(\omega)$, and $P(\omega)g(\omega)$ for $16$ number of twists when the $\epsilon$ values of LJ interactions (in kcal/mol) change while the $\sigma$ values are kept constant as $3.95$ \AA\ .}
\label{dos-inter}
\end{figure*}

Figure \ref{dos-inter} shows the resulting density of states, $g(\omega)$, and $P(\omega)g(\omega)$ for the aforementioned well depth magnitudes when the number of twists is $16$. As the well depth $\epsilon$ increases, the density of low frequency modes decreases while that of higher frequency modes increases. In particular, for low frequency modes less than $250$ cm$^{-1}$, the magnitude of  $P(\omega)g(\omega)$ generally decrease as the well depth magnitude increase.
However, we find that (generally in contrast to the work by Feldman {\it et.  al.}\cite{participationRatio1} on the low $g(\omega)$ region in their harmonic theory) trends in  $P(\omega)g(\omega)$ are not clearly correlated with the computed thermal conduction for the systems we study here. For instance, when the well depth increases from $0.0912$ kcal/mol to $0.2$ kcal/mol, we observe a decrease in $P(\omega)g(\omega)$ at low frequencies as the thermal conductance decreases. 
However, when the well depth increases from $0.00001$  kcal/mol to $0.0912$ kcal/mol, $P(\omega)g(\omega)$ at low frequencies decreases while the thermal conductance increases. Here we find that the smoothness of the helices is well correlated with the thermal conductance unlike  $P(\omega)g(\omega)$.
This may be attributed to the fact that the relationship between the $P(\omega)g(\omega)$ and thermal conductance is not direct and other factors such as phonon group velocities, mean free paths, and scattering processes also play crucial roles in determining thermal conductance.\cite{participationRatio1,participationRatio2}

\clearpage

\begin{figure*}[tbh]
\includegraphics[width=0.9\linewidth]{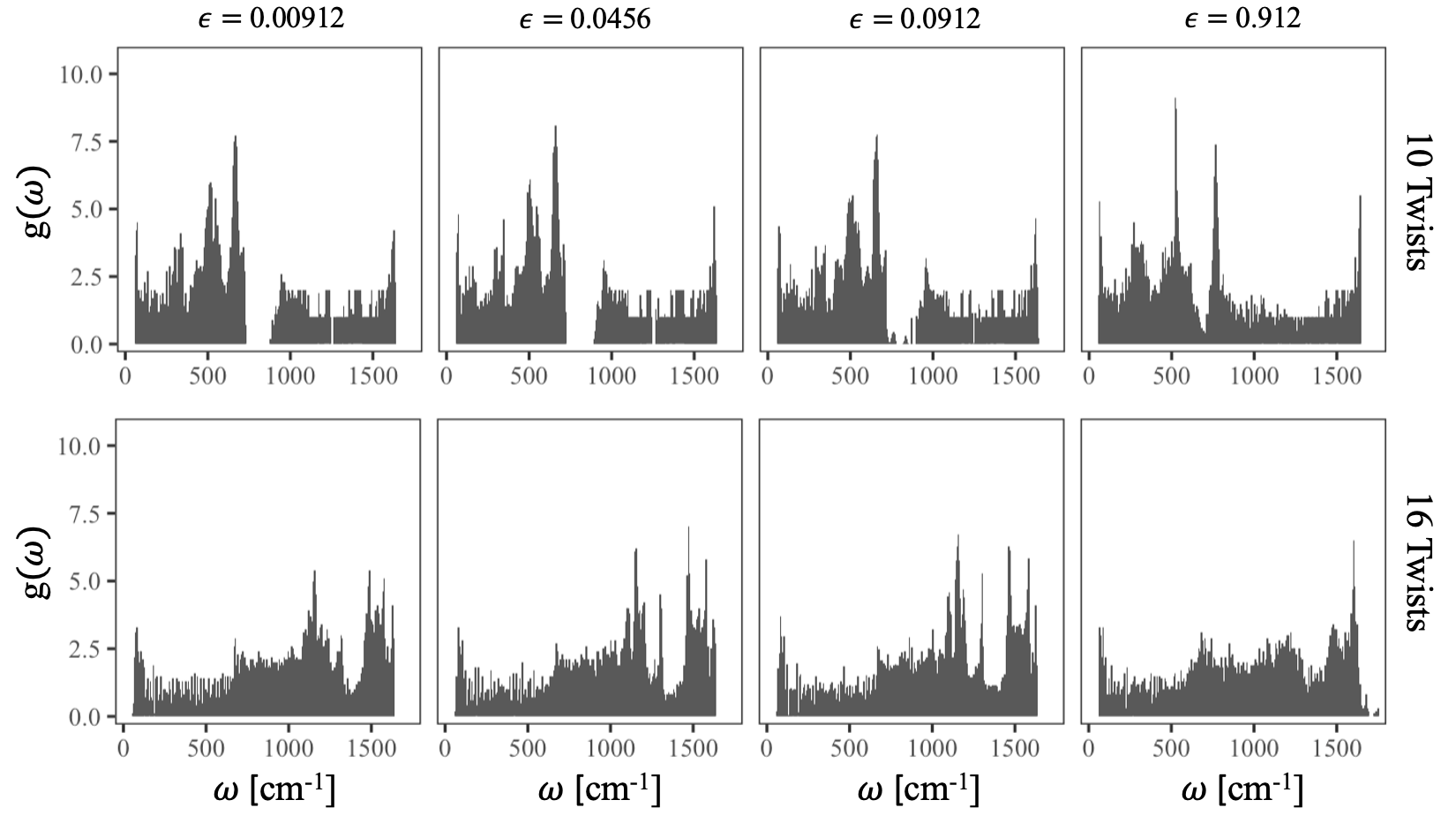}
\caption{The spectral densities of modes, $g(\omega)$  for $10$ and $16$ number of twists when the $\epsilon$ values of LJ interactions (in kcal/mol) change while the $\sigma$ values are kept constant as $3.95$ \AA\ .}
\label{dos_10_16}
\end{figure*}

To further investigate the influence of intermolecular interactions on the density of states ($g(\omega)$), we calculate  $g(\omega)$ for two polymer chains with $10$ and $16$ number of twists when the well depth varies. The top panel of Fig. \ref{dos_10_16} shows that when the number of twists is $10,$ there is a gap in the middle of frequency range of $g(\omega)$ that disappears as well depth magnitude increases. When number of twists is $16$, however, the gaps disappears in the middle frequency range for all well depth magnitudes examined. Therefore, it appears that increased inter-chain interactions diminish the gap in the middle of the frequency range of $g(\omega)$, since increase in the number of twists as well as the increase in well depth magnitudes are both factors that tend to diminish the gap.

\clearpage

Table \ref{inter_sigma} also shows how the thermal conductance for 2 polymer chains with $16$ number of twists change as $\sigma$ in the LJ parameters changes and non-monotonic behavior is observed. 
We do not go to further analyses here because we find the linear change in the well depth magnitudes is a more straightforward control for the intermolecular interactions between the polymer chains.

\begin{table}[tbh]
\centering
\caption{Thermal conductance values, $K$ , for 2 polymer chains with $16$ number of twists as the intermolecular interactions change.  The intermolecular potential is the LJ potential with $\epsilon$ as the well depth and $\sigma$ as the parameter for the hard sphere diameter. Here the $\epsilon$ values are kept constant to a value of $0.0912$ kcal mol$^{-1}$ while the $\sigma$ values change.}
\begin{tabular}{|c|c|}
     \hline
      $\sigma$, \AA\ & $K$, pW/K \\
     \hline
  $2.95$ & $1568\pm60$  \\
  $3.45$ & $2294\pm11$  \\
  $3.95$ & $2315\pm20$  \\
  $4.45$ & $1318\pm11$  \\
  $4.95$ & $797\pm10$  \\
     \hline
\end{tabular}
\label{inter_sigma}
\end{table}

\clearpage

\section{Temperature Dependence of Thermal Conductance}
Here we systematically investigate the temperature dependencies of heat conductance for two cases. First, we show how the heat conductance values change with zero number of twists for two polymer chains when the temperature changes from $300$ K to $500$ K
in the intervals of $50$ K.

\begin{table}[tbh]
\centering
\caption{The temperature dependencies of thermal conductance values for two PE polymer chains with chain lengths of $98$ for untwisted chains.}
\begin{tabular}{|c|c|}
     \hline
     $T$, K & $K$, pW/K \\
     \hline
     $300$ & $738\pm6$  \\
     $350$ & $670\pm4$  \\
     $400$ & $597\pm12$  \\
     $450$ & $550\pm8$  \\
     $500$ & $500\pm5$  \\
     \hline
\end{tabular}
\label{temperature}
\end{table}

Table \ref{temperature} shows that as the temperature increases, the heat conductance decreases for the junctions we study. 
This may be attributed to more jiggling of the atoms in the polymers leading to more scattering events.
Our observation is consistent with the results from Muthaiah,\cite{Muthaiah-2018} where they observed that the thermal conductance of stretched bulk polymer chains decreases from $200$ K to $350$ K. It is worth noting that for unoriented polyethylene, Luo {\it et. al.}\cite{ZhangTeng-2016} observed that as the temperature increases, the thermal conductivity initially increases, peaks around 350 K, and then decreases.

\begin{figure*}[tbh]
\includegraphics[width=0.6\linewidth]{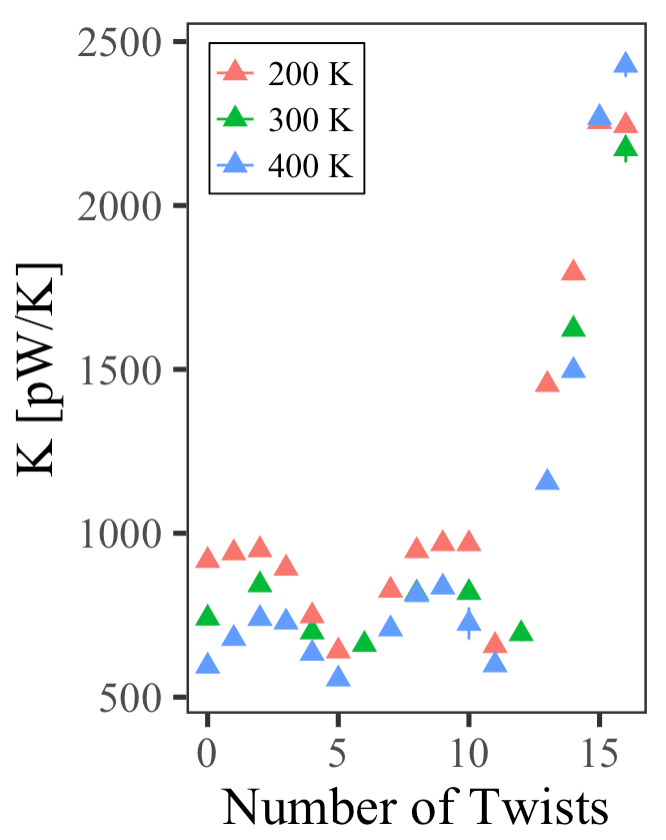}
\caption{The temperature dependencies of thermal conductance values from classical MD simulations as the number of twists change for two polymer chains.}
\label{temp}
\end{figure*}

Figure \ref{temp} shows how the thermal conductance values change for two PE polymer chains with length $N=98$ as the number of twists changes for several average temperatures. Each color corresponds to a different average temperature of the polymer wire, where the two Langevin regions differ by 40 K. Overall, we see that the primary qualitative trends of this study persist across various temperature. Generally, for most of the twist numbers, the thermal conductance values decrease as the temperature increases. Only in the high-twist regime does this trend tend to deviate. It should be noted that a good understanding of the temperature dependencies of the thermal conductance requires a rigerous consideration of Bose-Einstein statistic, which awaits future work.\cite{Renai-2023}

\clearpage

\section{Further Discussion of Polymer Wires with More than Two Chains}
\label{appendix:manyChains}

\begin{figure*}[h]
\includegraphics[width=0.9\linewidth]{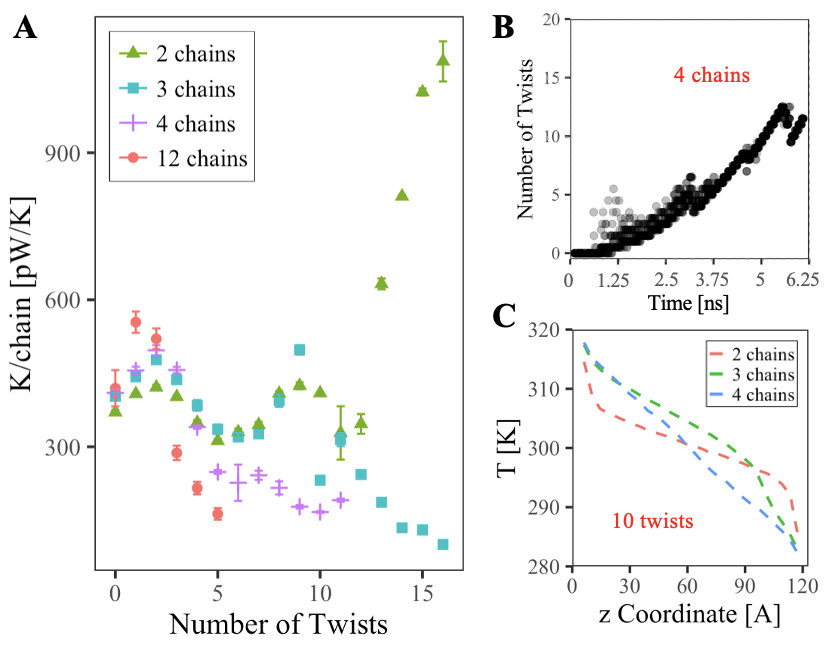}
\caption{\textbf{(A)} The thermal conductance per chain plotted as a function of the number of twists for wires consisting of 2 (green triangles), 3 (blue squares), 4 (purple crosses), and 12 (red circles) polyethylene chains of $N=98$. The part of this figure up to 5 twists is the same as \hyperref[FIG5]{Fig. 5.} \textbf{(B)} Number of twists in a polyethylene wire composed of 4 chains throughout the equilibration of MD simulations to generate twists. The number of twists for each point in this figure is computed naively based on the number of times the chains crossed, leading to some baseline fluctuations in the measurement. \textbf{(C)} Temperature profiles obtained during the production period of MD simulations for wires composed of 2 (red), 3 (green), and 4 (blue) polyethylene chains containing 10 twists during simulations used to compute (A).} 
\label{FIG14}
\end{figure*}

In the main text, we have included only results for $\leq5$ twists for polymer wires composed of more than 2 strands. This is because the concept of twist saturation was more difficult to understand for thicker wires. For example, in \hyperref[FIG2]{Fig. 3A} we saw that when torque was applied to the two-stranded wire, its twist grew continuously and plateaued at twist-saturation due to bond breakages. \hyperref[FIG13]{Figure \ref{FIG13}B} shows that this is not the case for the 4-stranded wire, where there are multiple discontinuities in the twist increase over time. We believe this is because when more than two chains are present, chains can cross each other if bonds can break and reform. This can result in non-trivial weaving patterns more conducive to twist increase. Furthermore, we think that the way the non-reactive force fields (used in this study) model these bond breakages and reformations might not be physically accurate because the atoms which are bonded to each other are specified by the input topology file. In reality, when an atom's bond breaks, it should be able to form a new bond with a different atom if the energy of that interaction is lowest. Here we report the results of these calculations. \hyperref[FIG13]{Figure \ref{FIG13}A} shows such thermal conductance results, and \hyperref[FIG13]{Figure \ref{FIG13}C} shows some corresponding temperature profiles. Note, for instance, that 10 twists could not be obtained in a wire that was 12 strands thick due to twist saturation and bond breakages.

\clearpage
\section{Details of the Explicit Hydrogen Model}
\label{appendix:withH}
For the Explicit Hydrogen model used in this work, the initial file containing the atomic coordinates was first generated artificially from typical bond lengths and angles. The structure was then allowed to relax at constant temperature of 300K using the described force field, thereby generating a realistic structure of a polymer wire. 

The force field parameters used in the Explicit Hydrogen model were taken from the literature as appropriate and are given in the following table. Equilibrium bonds lengths and angles as well as angular spring constants were taken from the standard TraPPE Explicit Hydrogen (EH) Force Field.\cite{FF4} Linear spring constants were computed from the accepted frequencies of organic bonds. Leonard Jones parameters were taken from the OPLS All-Atom Force Field.\cite{FF2}

\begin{table}[H]
\begin{center}
\caption {Explicit Hydrogen force field used in this work.} \label{tab:title} 
\begin{tabular}{ccc}
     \hline \hline
     \multicolumn{2}{c}{Bond potentials: $U_{\rm{bond}} =  k_b (l-l_0)^2$ }  \\
       & $k_b$, kcal mol$^{-1}$ \AA$^{-2}$ & $l_0$, \AA \\
      CH$_{x}$--CH$_{y}$  & $450$ &  $1.54$ \\
      CH$_{x}$--H  & $358$ &  $1.10$ \\
     \hline \hline 
     \multicolumn{2}{c}{Angle potential: $U_{\rm{angle}} =  k_\theta (\theta-\theta_0)^2$ }  \\
       & $k_\theta$, kcal mol$^{-1}$ rad$^{-2}$ & $\theta_0$, deg \\
     CH$_{x}$--CH$_{y}$--CH$_{z}$  & $58.4$ & $112.7$ \\
     CH$_{x}$--CH$_{y}$--H  & $39.38$ & $110.7$ \\
     H--CH$_{x}$--H  & $40.54$ & $107.8$ \\
     \hline \hline
     \multicolumn{2}{c}{Dihedral potential: $U_{\rm{dih}} =  \Sigma_n^4 \frac{C_n}{2} \left[ 1+ (-1)^{n-1} {\rm{cos}} ( n \phi) \right]$ }  \\
       & $C_i$, kcal mol$^{-1}$ & $C_i$ \\
      & $1.740$  & $C_1$ \\
     CH$_{x}$--CH$_{y}$--CH$_{z}$--CH$_{w}$  & $-0.157$ & $C_2$ \\
      & $0.279$   & $C_3$ \\
       & $0$   & $C_4$ \\ 
       \hline
      & $0$  & $C_1$ \\
     CH$_{x}$--CH$_{y}$--CH$_{z}$--H  & $0$ & $C_2$ \\
      & $0.366$   & $C_3$ \\
       & $0$   & $C_4$ \\
       \hline
       & $0$ & $C_1$ \\
     H--CH$_{x}$--CH$_{y}$--H  & $0$ & $C_2$ \\
      & $0.318$   & $C_3$ \\
       & $0$   & $C_4$ \\
     \hline \hline
     \multicolumn{2}{c}{Non-bonded potential: $U_{\rm{LJ}} =  4\epsilon \left[ (\sigma/r)^{12} - (\sigma/r)^6 \right] $ }  \\
       & $\epsilon$, kcal mol$^{-1}$ & $\sigma$, \AA \\
      C,~C (sp$^3$) & $0.066$ & $3.5$ \\
      C,~H (sp$^3$) & $0.044$ & $3.0$ \\
      H,~H (sp$^3$) & $0.030$ & $2.5$ \\
     \hline \hline
\end{tabular}
\end{center}
\end{table}

\newpage
\pagebreak

\section{Mathematical Details of the Continuous Chirality Measure (CCM)}
\label{appendix:ccm}
Let $Q=\ket{q_1},...,\ket{q_N}$ be a molecular structure where $\ket{q_k}$ is the vector representing the atomic coordinate of the $k$-th atom with respect to the center of mass. In this paper, we define the CCM of the molecule $Q$ as \begin{equation}\tag{5}
CCM(Q) = 1 - \frac{\sum_{k = 1}^{N} \bra{q_k}\hat{\mathds{1}}+\hat{\sigma_n}\ket{q_k}}{2\sum_{k = 1}^{N} \braket{q_k | q_k}},
\end{equation}
where $\sigma_n$ is the reflection operator over the plane defined by the optimal unit normal $\ket{n}.$ The optimal unit normal $\ket{n}$, which defines the plane of reflection that minimizes the CCM, can be found using Lagrange multipliers.\cite{csm} In the mathematical literature, the CCM typically optimizes over all permutations $\ket{p_k} = \hat{P}\ket{q_k}$ in addition to optimizing the plane of reflection. However, in the spirit of chemistry, we force the trivial permutation in order to honor the information contained in the molecule's bond connectivity. We have checked that for the helical structures studied in this work, the trivial permutation yields the lowest CCM value.

Note that the CCM can be understood intuitively as a structure's distance from its nearest achiral structure, where $\frac{1}{2} \Sigma_k(\hat{\mathds{1}}+\hat{\sigma_n})\ket{q_k}$ represents the closest achiral structure to $Q$ (i.e. the average of the two enantiomers). This appendix is not intended to be a thorough treatment of the CCM. Readers who have further interest should consult Ref. \citen{c1b, c2}.

\section*{References}
\bibliographystyle{aipnum4-1}

\end{document}